\documentclass[aps,prd,twocolumn,nofootinbib,floatfix,showpacs]{revtex4}
\usepackage[latin1]{inputenc}  
\usepackage{bbm,slashed}
\usepackage{mathrsfs}
\usepackage{graphicx,epsfig}
\usepackage{amsmath,amsfonts,amssymb}
\usepackage{booktabs}

\def\Z{\mathbbm{ Z}} 
\def\R{\mathbbm{R}}
\newcommand{\fdi}{\slashed{\partial}}
\newcommand{\cQ}{\mathcal{Q}}
\newcommand{\gam}{\gamma}
\newcommand{\gams}{\gamma_*}
\newcommand{\psib}{\bar{\psi}}

\newcommand{\thetab}{\bar{\theta}}
\newcommand{\pa}{\partial}

\newcommand{\eps}{\epsilon}
\newcommand{\epsb}{\bar\epsilon}

\newcommand{\ft}[2]{{\textstyle\frac{#1}{#2}}}
\newcommand{\ha}{\frac{1}{2}}

\newcommand{\al}{\alpha}
\newcommand{\chib}{\bar\chi}

\newcommand{\W}{\mathscr W}
\newcommand{\abar}{a}
\newcommand{\lambar}{{\lambda}}
\newcommand{\bbar}{b}
\newcommand{\w}{\mathfrak w}
\newcommand{\ua}{{ u}}

\newcommand{\ufrak}{\mathfrak u}

\graphicspath{{plots/}}

\DeclareMathOperator{\tr}{tr}
\DeclareMathOperator{\STr}{STr}
\DeclareMathOperator{\Tr}{Tr}

 \DeclareMathAlphabet{\boldmathe}{T1}{cmr}{bx}{it}
\newcommand{\abs}[1]{\left| #1 \right|}

\newcommand{\cD}{{\mathcal D}}

\newcommand{\id}{\mathbbm{1}}
\newcommand{\mtxt}[1]{\quad\hbox{{#1}}\quad}

\newcommand{\lam}{\lambda}

\def\G{\Gamma}

\def\t{\theta}
\def\tb{\bar{\theta}}
\def\pr{\prime}

\graphicspath{{plots/}}

\begin{document}
\title{Phase Diagram and Fixed-Point Structure of two dimensional $\mathcal N=1$
Wess-Zumino Models
}
\author{Franziska Synatschke, Holger Gies and Andreas Wipf}
\affiliation{
Theoretisch-Physikalisches Institut,\,\,Friedrich-Schiller-Universit{\"a}t
Jena,
Max-Wien-Platz~1, D-07743~Jena, Germany}

%======================================================
\begin{abstract}
We study the phases and fixed-point structure of two-dimensional
supersymmetric Wess-Zumino models with one supercharge. Our work is based on
the functional renormalization group formulated in terms of a manifestly
off-shell supersymmetric flow equation for the effective action. Within the
derivative expansion, we solve the flow of the superpotential also including
the anomalous dimension of the superfield. 
The models exhibit a surprisingly rich fixed-point structure with a discrete
number of fixed-point superpotentials. Each fixed-point superpotential is
characterized by its number of nodes and by the number of RG relevant
directions. In limiting cases, we find periodic superpotentials and potentials
which confine the fields to a compact target space. The maximally
IR-attractive fixed point has one relevant direction, the tuning of which
distinguishes between supersymmetric and broken phases. For the Wess-Zumino
model defined near the Gau\ss ian fixed point, we determine the phase diagram
and compute the corresponding ground-state masses.
\end{abstract}
%======================================================
\maketitle

%======================================================
\section{Introduction}
%======================================================

Supersymmetry has become a well-received guiding principle in the construction
of particle-physics models beyond the standard model. Whereas the resulting
phenomenology of these models is often worked out by perturbative analysis,
the required breaking of supersymmetry may be of nonperturbative origin; it is
therefore typically parameterized into the models by a phenomenological
reasoning. For a proper understanding of the underlying dynamical mechanisms
of symmetry breaking which is often related to collective condensation
phenomena, powerful and flexible nonperturbative methods specifically adapted
to supersymmetric theories will eventually be needed.

For many nonperturbative problems in field theory, lattice formulations and
simulations have proven successful. As supersymmetry intertwines field
transformations with spacetime translations, discretizing spacetime often
induces a partial loss of supersymmetry. This problem also goes along with the
challenge of properly implementing dynamical fermions on the lattice,
currently witnessing significant progress 
\cite{Catterall:2009it,Giedt:2006pd,Bergner:2007pu,Kastner:2008zc}. As completely
new territory is entered in these studies, nonperturbative continuum methods
which can preserve supersymmetry manifestly can complement the lattice
studies, eventually leading to a coherent picture. 

A promising candidate for a nonperturbative method is the functional
renormalization group (RG) which has been successfully applied to a wide range
of nonperturbative problems such as critical phenomena, fermionic systems,
gauge theories and quantum gravity, see 
\cite{Aoki:2000wm,Berges:2000ew,Litim:1998nf,Pawlowski:2005xe,Gies:2006wv,Sonoda:2007av}
for reviews. A number of conceptual studies of supersymmetric theories has
already been performed with the functional RG. The delicate point here is, of
course, the construction and use of a manifestly supersymmetry-preserving
regulator. For instance, a supersymmetric regulator for the four-dimensional
Wess-Zumino model has been presented in \cite{Vian:1998kv,Bonini:1998ec}. A functional RG
formulation of supersymmetric Yang-Mills theory employing the superfield
formalism has been given in \cite{Falkenberg:1998bg}; for applications, see
also \cite{Arnone:2004ey,Arnone:2004ek}. Recently, general theories of a
scalar superfield including the Wess-Zumino model have been investigated with
a Polchinski-type RG equation in \cite{Rosten:2008ih}, yielding a new approach
to supersymmetric nonrenormalization theorems. A Wilsonian effective action
for the Wess-Zumino model by perturbatively iterating the functional RG has
been constructed in \cite{Sonoda:2008dz}.

This work is devoted to the two-dimensional $\mathcal N=1$ Wess-Zumino model
with a general superpotential, exploring the model beyond the realm of
perturbative expansions around zero coupling. This is the simplest quantum
field theoretic and supersymmetric model where the nonperturbative dynamical
aspects of supersymmetry breaking can be studied. The present study details
and generalizes our results presented in a recent Letter \cite{Gies:2009az},
and builds on our earlier work on supersymmetric quantum mechanics, where we
have constructed a manifestly supersymmetric functional RG flow for the
anharmonic oscillator \cite{Synatschke:2008pv}; see also
\cite{Horikoshi:1998sw,Weyrauch:2006aj} for RG studies of supersymmetric
quantum mechanics.

Inspired by Witten's work on the potential breaking of supersymmetry in the
Wess-Zumino model \cite{Witten:1982df}, pioneering nonperturbative lattice
studies based on Hamiltonian Monte-Carlo methods had early been performed for
this model by Ranft and Schiller \cite{Ranft1984166}.  More recently Beccaria and coworkers
\cite{Beccaria:2004ds,Beccaria:2004pa} re-investigated the phase diagram and
the ground-state energy of the model with similar methods. Golterman and Petcher
\cite{Golterman:1988ta} formulated a lattice action with a partially realized
supersymmetry. Another lattice
study of the Wess-Zumino model has been performed by Catterall and Karamov
\cite{Catterall:2003ae}. The results of these studies give a first glimpse
into the properties of the phase diagram as will be discussed in
Sect. \ref{sec:gaussian-fixed-point}. 

In the present paper, we use the functional RG equation for the superpotential
$W$ to study the phase structure of the ${\cal N}=1$ Wess-Zumino model in two
dimensions.  In Sect.~\ref{sec:results-from-one}, we begin with recalling the
off-shell and on-shell effective potentials in a one-loop approximation and
comment on certain flaws of the approximations. The main part of this work is
concerned with extending and applying the manifestly supersymmetric RG
techniques developed in \cite{Synatschke:2008pv} in the context of
supersymmetric quantum mechanics. The manifestly supersymmetric flow equation
for the effective action is constructed in Sect.~\ref{sec:superRG}. To first
order in a derivative expansion of the effective action, we solve the RG flow
equation for the effective superpotential in
Sect.~\ref{sec:local-potent-appr}. 

In the fixed-point analysis performed in Sect.~\ref{sec:FixedPointStructure},
already a simple polynomial expansion of the superpotential gives access to an
infinite number of fixed points with an increasing number of RG relevant
directions. Beyond the polynomial expansion, the flow equation for the full
superpotential reveals a variety of qualitatively different solutions
depending on the initial conditions: we find periodic, sine-Gordon type
solutions as well as sigma-model type solutions confining the field values to
a finite interval. At next-to-leading order, a nonzero anomalous dimension
governs the large-field asymptotics of the fixed-point superpotentials, such
that a family of regular fixed-point solutions arises. This family of
superpotentials shows oscillating behavior for small fields and a standard
asymptotics for large fields, similar to fixed-point potentials for pure
bosonic theories in two dimensions \cite{Morris:1994jc,Neves:1998tg}. As a
particularity of these supersymmetric models, we identify a new scaling relation
between the leading critical exponent of the superpotential flow and the
anomalous dimension.

Finally, we study the phase diagram of a particular Wess-Zumino model defined
near the Gau\ss ian fixed point in terms of a quadratic superpotential
perturbation in Sect.~\ref{sec:gaussian-fixed-point}. Depending on the initial
values of the control parameters of the potential, we observe a quantum phase
transition from the supersymmetric to the dynamically broken phase. Following
the lattice studies \cite{Beccaria:2003ba,Beccaria:2004ds,Beccaria:2004pa}, we
calculate the critical value of the control parameter for the phase transition
as a function of the coupling $\lambda$ at
the cutoff $\Lambda$, but now for all values of the coupling $\lambda$. We also
determine the fermionic and bosonic masses in both phases.

%======================================================
\section{Wess-Zumino model}
%======================================================

Two-dimensional Wess-Zumino models with one supersymmetry are 
particular Yukawa models where the self-interaction of the scalar 
field determines the Yukawa coupling. In an off-shell formulation 
they contain a scalar field $\phi$,  a Majorana spinor field $\psi$ and an
auxiliary field $F$. To maintain supersymmetry in every step of our 
calculations we combine these fields to one \emph{real superfield}
\begin{equation}
\Phi(x,\theta)=\phi(x) +\tb\gams\psi(x) 
+\ha(\tb\gams\theta) F(x).\label{model3}
\end{equation}
The anticommuting $\theta$ parameter in this
expansion is a constant Majorana spinor.
Supersymmetry transformations are generated by the supercharges
\begin{equation}
 \cQ=-i\frac{\partial}{\partial\tb}-\fdi\t,\quad
\bar\cQ=-i\frac{\partial}{\partial\t}-\tb\fdi,\label{model5}
\end{equation}
which anticommute on space-time translations,
$\{\cQ_\al,\bar \cQ_\beta\}=2i\fdi_{\al\beta}$.
The transformation rules for the component fields are obtained
by comparing coefficients in $\delta \Phi=i\epsb[\cQ,\Phi]$ and read
\begin{equation}
  \begin{split}
\delta \phi=\epsb\gam_*\psi,\quad
\delta\psi=(F+i\gam_*\fdi\phi)\eps,\\
\delta\psib=\epsb(F-i\fdi\phi\gam_*),\quad
\delta F=i\epsb\fdi\psi.\label{model11}
\end{split}
\end{equation}
As usual the $F$ term transforms into a total derivative such that
its space-time integral is invariant under
supersymmetry transformations. The supercharges anticommute with 
the superderivatives
\begin{align}
\cD=\frac{\partial}{\partial\tb}+i\fdi\t,\quad
\bar\cD=-\frac{\partial}{\partial\t}-i\tb\fdi,
\label{model7}
\end{align}
and up to a sign they obey the same anticommutation 
rules as the supercharges,
$\{\cD_\al,\bar \cD_\beta\}=-2i\fdi_{\al\beta}$.
In explicit calculations, one uses
Fierz identities which all follow from
\begin{equation}
\psi\chib=-\ft12\chib\psi-\ft12\gam_\mu(\chib\gam_\mu\psi)-\ft12\gam_*(\chib\gam_*\psi),
\label{model9}
\end{equation}
where $\gams=i\gam_0\gam_1$ anti-commutes with the $\gam_\mu$.
It is useful to keep in mind that for Majorana spinors the 
fermionic bilinears have the symmetry properties
\begin{align}
\psib\chi=-\chib\psi,\; \psib\gam_\mu\chi=-\chib\gam_\mu\psi
\mtxt{and}
\psib\gams\chi=\chib\gams\psi,\label{model1}
\end{align}
such that the only Lorentz-invariant bilinear is $\psib\gam_*\psi$ since
$\psib\psi=0$. Thus, we choose as Lagrangian density ${\cal L}_0$ for the free
theory the $D$ term of
\begin{equation}
\ft12\bar \cD\Phi\gam_* \cD\Phi
=\ft12\psib\gam_*\psi+(\thetab\gam_*\psi)F-i\thetab\gam^\mu\psi\pa_\mu \phi-
\thetab
\gam_*\theta{\cal L}_0.\label{model13}
\end{equation}
In components it has the form
\begin{equation}
{\cal L}_0=\ft12\pa_\mu \phi\pa^\mu \phi
+\ft{i}{4}\psib\fdi\psi-\ft{i}{4}\pa_\mu\psib\gam^\mu\psi
-\ft12 F^2.\label{model15}
\end{equation}
In this work, we study a class of interacting theories, where the interaction
Lagrangian is given by the $D$ term of a superpotential $W(\Phi)$,
\begin{equation}
W(\Phi)=W(\phi)+\thetab\gam_*\psi W'(\phi)-\ft12\thetab\gam_*\theta 
{\cal L}_1.\label{model17}
\end{equation}
In components, ${\cal L}_1$ has the form
\begin{equation}
{\cal L}_1=\ft12 W''(\phi)\psib\gam_*\psi-W^\pr(\phi)F.\label{model19}
\end{equation}
The sum of ${\cal L}_0$ and ${\cal L}_1$ defines the \emph{off-shell Lagrangian density}
\begin{equation}
{\cal L}
=\ft12\pa_\mu \phi\pa^\mu \phi
+\ft{i}{2}\psib\fdi\psi-\ft12 F^2
+\ft12 W''(\phi)\psib\gam_*\psi-W^\pr(\phi)F,\label{model21}
\end{equation}
which gives rise to an invariant action. As expected for a Euclidean model,
this action is unbounded from below and above. After eliminating the auxiliary
field via its algebraic equation of motion
\begin{equation}
F=-W^\pr(\phi),\label{model23}
\end{equation}
we end up with the stable \emph{on-shell Lagrangian density}
\begin{equation}
{\cal L}=\ft12(\pa \phi)^2+\ft{i}{2}\psib\fdi\psi+\ft12 W'{\,^2}(\phi)
+\ft12 W''(\phi)\,\psib\gam_*\psi.\label{model25}
\end{equation}
This density is invariant under the nonlinear on-shell
supersymmetry transformations
\begin{equation}
  \begin{split}
\delta \phi=\epsb\gam_*\psi,\quad
\delta \psi=\big(i\gam_*\fdi \phi-W'(\phi)\big)\eps,\\
\delta\psib=\epsb\big(i\gam_*\fdi \phi-W'(\phi)\big).\label{model27}
\end{split}
\end{equation}
For a polynomial superpotential $W$ the supersymmetric Yukawa 
models defined in Eq.~\eqref{model21} and \eqref{model25} are
perturbatively super-renormalizable.
If the leading term in the superpotential contains an even power of $\phi$,
$W=c\phi^{2n}+\mathcal O(\phi^{2n-1})$,  supersymmetry cannot be broken
by quantum corrections. However, if the leading term contains an odd power,
 supersymmetry may be broken.

%======================================================
\section{One-loop perturbation theory}
%======================================================
\label{sec:results-from-one}

Let us first discuss the effective potential in one-loop approximation, which
can be set up in both the on-shell or the off-shell formulation. It is
well-known that the one-loop on-shell potential becomes artificially complex
for nonconvex classical potentials \cite{Coleman:1973jx}. This problem is
avoided in the off-shell one-loop formulation: here, keeping first the
auxiliary field in the one-loop calculation and subsequently eliminating it by
its quantum equation of motion corresponds to a resummation of higher-order
terms in the on-shell formulation.  We expect that this potential is a better
approximation to the exact effective potential as compared to the one-loop
on-shell potential. Indeed, we find a real and stable effective potential
based on the off-shell calculation. Similar observations can be found in
\cite{Murphy:1983ag}.  The corresponding problem in one-dimensional
supersymmetric systems has been carefully analyzed by Bergner (see
\cite{Bergner:2009} and references therein).

\subsection{On-shell effective potential}

To calculate the one-loop potential in the on-shell
formulation, we need the fluctuation operators $M_{\mathrm{F}}$ and $M_{\mathrm{B}}$
for the fermion and the remaining scalar, respectively. For a homogeneous
background field $\phi$ playing the role of a mean field, these operators read
\begin{equation}
  \begin{split}
M_{\mathrm{B}}=p^2+V''\mtxt{and}
M_{\mathrm{F}}=i\fdi+\gam_*W''\\\Rightarrow
M_{\mathrm{F}}^2=(p^2+W''^2)\id_2,
\label{loop1}
  \end{split}
\end{equation}
where $V=\ft12 W'^{\,2}$ is the classical potential for the scalar
field. In a finite 
box of size $L$, the momentum takes the values $p_\mu=2\pi n_\mu/L$
with integer-valued $n_\mu$. The fermionic
integration yields the Pfaffian of $M_{\mathrm{F}}$,
\begin{equation}
\hbox{Pf}(M_{\mathrm{F}})=\pm \sqrt{\det(M_{\mathrm{F}})}.\label{loop3}
\end{equation}
We shall assume that the Pfaffian has a fixed sign, such that we may replace
the Pfaffian by the square root of the determinant. In a perturbative approach
and in the broken phase (with vanishing Witten index), this assumption is
justified.

For an operator $M$ with eigenvalues $p^2+C^2$ 
the derivative of the
zeta function $\zeta_M(s)=\tr (M/\mu^2)^{-s}$ at the origin is
\begin{equation}
\zeta_M'(0)=\frac{(CL)^2}{4\pi}\left(\ln\left(\frac{C}{\mu}\right)^2-1+4\sum_{
n_\mu\neq 0}\frac{K_1(CL n)}
{CL n}\right)\label{loop5}
\end{equation}
with the last sum, which involves the MacDonald function, approaching zero
exponentially fast with increasing box sizes. Thus, in the thermodynamic limit
the one-loop effective potential in the zeta-function scheme is
\begin{equation}
  \begin{split}
U^{(1)}_{\rm on}=&\ha W'^2+\frac{1}{2L^2}\left(\zeta'_{M_{\mathrm{F}}}(0)-\zeta'_{M_{\mathrm{B}}}(0)\right)
\\
=&\ha W'^2-\frac{W''^2}{8\pi}[(1+X)\ln\left(1+X\right)\\
&\phantom{\ha W'^2-\frac{W''^2}{8\pi}[}+X\ln(W''/\mu)^2-X
],\label{loop7}
\end{split}
\end{equation}
with $X=W'W'''/W''^2$. The energy scale $\mu$ is fixed by 
a renormalization condition. If we use a momentum cutoff
regularization instead of the $\zeta$-function regularization
then we obtain the same result with $\mu$
replaced by the cutoff $\Lambda$. 

The effective potential becomes complex for 
nonconvex classical potentials  $V=\ft12 W'^{\,2}$.
To be specific, let us choose 
\begin{equation}
W'=\bar\lambda(\phi^2-\bar a^2)\quad \Rightarrow\quad 
V=\frac{\bar\lam^2}{2}(\phi^2-\bar a^2)^2.
\label{loop9}
\end{equation}
For negative $\bar a^2$, the one-loop effective potential is real. For positive
$\bar a^2$, it becomes complex for small fields $\phi^2< \bar a^2/3$. For fields
slightly bigger than $\bar a/\sqrt{3}$, the potential is real and
\emph{negative}. This signals the failure of the approximation since the
effective potential must be non-negative in a supersymmetric theory. Depending
on the sign of the renormalized $ a^2$, we find both a supersymmetric phase
characterized by a non-vanishing expectation value
$\langle\phi\rangle=\phi_{\rm min}$ and $U^{(1)}_{\rm on}(\phi_{\rm min})=0$
and a phase with $\langle\phi\rangle=0$ and broken supersymmetry, see 
Fig.~\ref{fig:pot}. Here, the renormalization conditions are chosen such that the
minimum $\phi_{\text{min}}$ agrees with the renormalized value $a$ in the
supersymmetric phase. In the broken phase, we use a simple renormalization
condition by fixing all parameters at the cutoff, $\mu=\Lambda$. In both
cases, the renormalization condition could alternatively be formulated in
terms of Coleman-Weinberg renormalization conditions by fixing the curvature
of the potential at the minimum to the physical mass. 

%======================================================
\subsection{Off-shell effective potential}
%======================================================

In the off-shell formulation, the fluctuations of both scalars (including the
auxiliary field) and the fermion are taken into account.  The off-shell
Lagrangian (\ref{model21}) gives rise to the fluctuation operators in the
background of both a $\phi$ and an $F$ mean field,
\begin{equation}
M_{\mathrm{B}}=\begin{pmatrix}p^2-FW'''&-W''\cr -W''&-1\end{pmatrix}\mtxt{and}
M_{\rm F}=\slashed{p}+\gam_*W'',\label{loop11}
\end{equation}
where the mean fields $\phi$ (as the argument of $W''$ and $W'''$) and $F$ are
assumed to be homogeneous. It follows that the $\zeta$-function regularized
one-loop off-shell potential reads
\begin{equation}
  \begin{split}
U^{(1)}_{\rm off}=-\ha F^2-FW'
-\frac{W''^2}{8\pi}[(1+Y)\ln(1+Y)\\
+Y\ln (W''/\mu)^2-Y
]\label{loop13}
\end{split}
\end{equation}
with $Y=-FW'''/W''^2$. With a momentum cutoff regularization, we obtain the
same result with $\mu$ denoting the cutoff. To eliminate the auxiliary field
$F$, we must solve the transcendental gap equation
\begin{equation}
\pa_F U^{(1)}_{\rm off}=
-F-W'+\frac{W'''}{8\pi}\ln \frac{(W''^2-FW''')}{\mu^2}=0,\label{loop15}
\end{equation}
and insert the solution for $F$ back into $U^{(1)}_{\rm off}$.
For an arbitrary $\phi$, the gap equation  always has a real solution $F$ leading to 
a real effective potential \cite{Bartels:1983wm}. Concerning supersymmetry breaking,
we find the same qualitative result as with the on-shell calculation:
the sign of the renormalized $a^2$ determines the phase of the system.

\begin{figure}[ht]
	\includegraphics[width=.9\columnwidth]{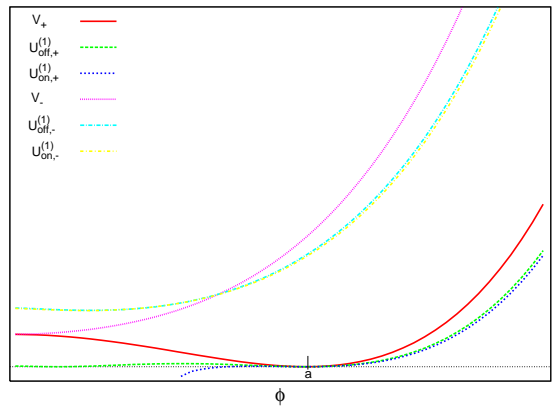} 
	\caption{The classical potential $V$
and the on-shell and off-shell effective potentials 
$U^{(1)}_{\rm on}$ and $U^{(1)}_{\rm off}$ in
one-loop approximation. The subscripts $+$ and $-$ denote the sign of $\bar
a^2$. }\label{fig:pot}
 \end{figure}

Note that the on-shell and off-shell potentials (\ref{loop7}) and
(\ref{loop13}) have similar forms. In Fig.~\ref{fig:pot} the two potentials
are compared with each other and also with the classical potential $V$. Of
course, the same renormalization conditions are used for the on- and off-shell
potentials. For positive $ a^2$, the scale parameter $\mu$ has been adjusted
such that the potentials take their minima at the same value $\phi_{\rm min}=
a$. %
Note that the off-shell potential contains resummed contributions from higher
order in $\hbar$. The reason is that the solution $F$ of (\ref{loop15})
contains terms of higher order in $\hbar$ and inserting the solution back into
the effective potential generates terms to all orders of $\hbar$ in the
effective potential. By expanding the off-shell action in $\hbar$, the
first-order result agrees again with the complex on-shell effective potential.
The effective resummation contained in the off-shell action is such that the
effective potential becomes real and non-negative everywhere, in particular,
at those points where the classical potential is not convex. For $ a^2>0$,
both one-loop potentials predict a phase with broken $\Z_2$ symmetry and unbroken
supersymmetry. For $ a^2<0$, we find a phase with unbroken $\Z_2$ symmetry
and broken supersymmetry.

%======================================================
\section{Supersymmetric RG flow}
%======================================================
\label{sec:superRG}

In this section, we will construct a manifestly supersymmetric flow equation in the
off-shell formulation. Our approach is based on the functional RG formulated in
terms of a flow equation for  the effective average action
$\Gamma_k$, i.e., the Wetterich equation,  
\cite{Wetterich:1992yh} 
\begin{equation}
 \partial_k\Gamma_k=
 \frac12 \STr\left\{\left[\Gamma_k^{(2)}+ R_k\right]^{-1}\partial_k  R_k\right\}.
\label{eq:lpa1}
\end{equation}
Here, $\Gamma_k$ is a scale-dependent effective action; it interpolates
between the microscopic or classical action $S$ for $k\to\Lambda$, with
$\Lambda$ being the microscopic UV scale, and the full quantum effective
action $\Gamma=\Gamma_{k\to0}$, being the standard generating functional for
1PI correlation functions. The interpolating scale $k$ denotes an infrared
IR regulator scale below which all fluctuations with momenta smaller than
$k$ are suppressed. For $k\to0$, all fluctuations are taken into account and
we arrive at the full solution of the quantum theory in terms of the effective
action $\Gamma$. The Wetterich equation defines an RG trajectory in the
space of action functionals with the classical action $S$ serving as initial
condition.

In Eq.~(\ref{eq:lpa1}), we encounter the second functional derivative of
$\Gamma_k$, 
\begin{equation}
\left(\Gamma_k^{(2)}\right)_{ab}=\frac{\overrightarrow{\delta}}{\delta\Psi_a}
\Gamma_k\frac {\overleftarrow{\delta}}{\delta\Psi_b}\,,\label{eq:lpa3}
\end{equation}
where the indices $a,b$ summarize field components, internal and Lorentz
indices, as well as spacetime or momentum coordinates. In the present case, we
have $\Psi^{\text T}=(\phi,F,\psi,\bar\psi)$ where $\Psi$ is not a superfield,
but merely a collection of fields. The momentum-dependent regulator function
$R_k$ in Eq.~(\ref{eq:lpa1}) establishes the IR suppression of modes below
$k$. In the general case, three properties of the regulator $R_k(p)$ are
essential: (i) $R_k(p)|_{p^2/k^2\to 0} >0$ which implements the IR
regularization, (ii) $R_k(p)|_{k^2/p^2\to 0} =0$ which guarantees that the
regulator vanishes for $k\to0$, (iii) $R_k(p)|_{k\to\Lambda\to\infty}\to
\infty$ which serves to fix the theory at the classical action in the UV.
Different functional forms of $R_k$ correspond to different RG trajectories
manifesting the RG scheme dependence, but the end point $\Gamma_{k\to 0}\to
\Gamma$ remains invariant.

The regularization preserves supersymmetry if the regulator contribution to
the action $\Delta S_K$ is supersymmetric, see below. As the regulator needs
to be quadratic in the fields in order to maintain the one-loop structure of
the flow, a general supersymmetric quadratic form can be constructed as a $D$
term of a superfield operator $\Phi K\Phi$. Here, $K$ is a function of the two
invariant and commuting operators $\bar \cD\gam_*\cD$ and $\bar \cD\fdi
\cD\propto (\bar \cD\gam_*\cD)^2$.  Since powers of $\bar \cD\gam_*\cD$
boil down to
\begin{equation}
  \begin{split}
(\ft12\bar \cD\gam_*\cD)^{2n}&=\ft{i}{2}\bar \cD\fdi
\cD(\partial^2)^{n-1}
\mtxt{and}
\\ 
(\ft12\bar \cD\gam_*\cD)^{2n+1}&=\ft12\bar
\cD\gam_*\cD(\partial^2)^n\label{eq:lpa7},
\end{split}
\end{equation}
where $\partial^2$ is the standard Laplacian, any invariant and quadratic
regulator action is the superspace integral of
\begin{equation}
\ha\Phi\bar \cD\big(\tilde r_1(-\partial^2)
-\gam_*r_2(-\partial^2)\big)\cD\Phi .\label{eq:lpa9} 
\end{equation}
Expressed in component fields, we find
\begin{equation}
\Delta S_k=\ha\int(\phi,F) R^{\rm B}_k\, {\phi \choose F}
+\ha\int \psib R^{\rm F}_k\psi.\label{eq:lpa11}
\end{equation}
In momentum space, $i\pa_\mu$ is replaced by $p_\mu$ and the operators take the
explicit form
\begin{equation}
R^{\rm B}_k=\begin{pmatrix}p^2 r_2 &-r_1\cr -r_1&-r_2\end{pmatrix}\mtxt{and}
R^{\rm F}_k=\slashed p\,r_2+\gam_*r_1,\label{eq:lpa13}
\end{equation}
where $r_1=p^2\tilde r_1$. Comparison with Eq.~(\ref{loop11}) reveals that
$r_1$ plays the role of a momentum-dependent supersymmetric mass term, whereas
$r_2$ can be viewed as a deformation of the momentum dependence of the kinetic
term.  

This choice of the regulator guarantees a supersymmetric RG trajectory;
i.e., for a supersymmetric initial condition $\Gamma_{k\to\Lambda}\to S$, the
solution to the flow equation $\Gamma_k$ will remain manifestly supersymmetric
for all $k$ including the endpoint $\Gamma=\Gamma_{k\to 0}$. This does not
only hold for the exact solution, but is also valid for truncated effective
actions, provided the truncation is built from supersymmetric field
operators.

%======================================================
\section{Local potential approximation}
%======================================================

\label{sec:local-potent-appr}

Various systematic and consistent approximation schemes for the construction
of $\Gamma_k$ can be devised with the flow equation. In this work, we use the
derivative expansion which is based on the underlying assumption that the
fully interacting theory remains sufficiently local if formulated in the given
set of field variables. In order to preserve supersymmetry, we expand the
effective action in powers of super-covariant derivatives in the off-shell
formulation. This expansion allows for a systematic and unique classification
of all possible operators. A truncation of the effective action to a finite
derivative order leads to a closed set of equations for the expansion
parameters. 

In this section, we concentrate on the leading-order derivative expansion: the
so-called local potential approximation. Here, the truncated effective
Lagrangian is given by Eq.~(\ref{model21}) with a scale-dependent
superpotential $W_k$, such that the truncated effective action reads
\begin{equation}
  \begin{split}
  \Gamma_k[\phi,F,{\psib},\psi]
 = \int d^2x
  \left(\ft12\partial_\mu\phi\partial^\mu\phi+
\ft{i}{2}\psib\slashed{\partial}\psi
	-\ft12F^2
	\right.\\+
	\left.	\ft12W_k''(\phi)\psib\gamma_\ast\psi-W_k'(\phi)F\right).\label{eq:lpa5}
	\end{split}
 \end{equation}
The derivation of the flow equation for the superpotential parallels the
corresponding one for supersymmetric quantum mechanics given in a previous
work \cite{Synatschke:2008pv}. Within the approximation of constant mean fields, the
second functional derivative of the effective action plus regulator is
\begin{equation}
  \begin{split}
 \Gamma^{(2)}_k+R_k&=\begin{pmatrix}A&W_k''' e_1\otimes\psib\gam_*\cr
W_k'''\gam_*\psi\otimes e_1^T&B  \end{pmatrix},\\
e_1&=(1,\;0)^T,
\label{eq:lpa15}
\end{split}
\end{equation}
where the operators on the diagonal read
\begin{equation}
  \begin{split}
&A=
\begin{pmatrix}p^2(1+r_2)
-FW_k'''+\ha  W_k^{(4)}\psib\gam_*\psi
&-W_k''-r_1\cr
 -W_k''-r_1&-1-r_2
\end{pmatrix},\\
\quad 
&B=i(1+r_2)\slashed{p}+\gam_*(r_1+W_k'').\label{eq:lpa19}
\end{split}
\end{equation}
The inverse of the operator defined in Eq.~(\ref{eq:lpa15}) can be written as follows,
\begin{equation}
  \begin{split}
\frac{1}{\Gamma^{(2)}_k+R_k}&=
\begin{pmatrix}
G_{k}^{\rm BB}&G_{k}^{\rm BF}\\G_{k}^{\rm FB}&G_{k}^{\rm FF}
\end{pmatrix}\\
&=
\begin{pmatrix}A^{-1}&0\cr 0&B^{-1}\end{pmatrix}K
\begin{pmatrix}A^{-1}&0\cr 0&B^{-1}\end{pmatrix},\label{eq:lpa21}
\end{split}
\end{equation}
where we have abbreviated
\begin{widetext}
\begin{equation}
K=\begin{pmatrix}A+W_k'''^2(\psib\gam_*B^{-1}\gam_*\psi)\, e_1\otimes e_1^T
&-W_k''' e_1\otimes\psib\gam_*\cr
-W_k'''\gam_*\psi\otimes e_1^T&
B-\ha W_k'''^2(e_1^TA^{-1}e_1) (\psib\gam_*\psi)\gam_*  \end{pmatrix}.\label{eq:lpa23}
\end{equation}
\end{widetext}
In order to verify that Eq.~(\ref{eq:lpa21}) is the inverse of
Eq.~(\ref{eq:lpa15}), the Fierz identity $\psi\otimes\psib=-\ha
\gam_*(\psib\gam_*\psi)$ is useful.
This result is inserted into the flow equation (\ref{eq:lpa1}), 
which in component notation reads
\begin{align}
  \partial_k\Gamma_k=
\ha\Tr \left(\pa_kR^{\rm B}_k\,G_k^{\rm BB}\right)
 -\ha\Tr \left(\pa_k R^{\rm F}_k G_k^{\rm FF}\right).\label{flow3}
 \end{align}
The flow equation for $W'_k(\phi)$  is obtained by projecting both
sides of this equation onto the term linear in the auxiliary field.  This yields
\begin{multline}
	\partial_kW'_k(\phi)=-W_k'''\int \frac{d^2p}{4\pi^2}\left(
	\frac{(1+r_2)(W_k''+r_1)}{\Delta^2}
		\partial_kr_1\right.
		\\+\left.
	\frac{p^2(1+r_2)^2-(W_k''+r_1)^2 
}{2\Delta^2}
		\partial_kr_2\right),
		\label{eq:flow}		
\end{multline}
where we have introduced $\Delta=p^2(1+r_2)^2+(W_k''+r_1)^2$.  Integrating
with respect to $\phi$ and dropping an irrelevant constant leads to
\begin{align}
	\partial_kW_k(\phi)=&\frac12\int \frac{d^2p}{(2\pi)^2}
	\frac{(r_2+1)\partial_kr_1-(r_1+W_k''(\phi ))\partial_kr_2}
	{\Delta}.
		\label{eq:flow1}
\end{align}
This flow equation for the superpotential has exactly the same structure as
the corresponding flow equation in supersymmetric quantum mechanics
\cite{Synatschke:2008pv}. This is not surprising since supersymmetric quantum
mechanics can be obtained from this model through dimensional reduction. 

Here we are interested in superpotentials for which the map $\R\ni\phi\to
W'(\phi)\in \R$ has winding number zero as these potentials allow for
dynamical supersymmetry breaking.  For a polynomial $W$ this is the case if
$W'$ tends asymptotically to an even power, $W'(\phi)\sim
\mathrm{c}\,\phi^{2n}$. Then, the highest power of $W''(\phi)$ is odd. This
implies that the mass-like regulator $r_1$ does not screen but merely shift
possible zeroes of $W_k''$.  Thus we may set $r_1=0$ without spoiling the IR
properties of the flow.

In the present local-potential approximation, the simple cutoff function
$r_2=(k/|p|-1)\,\theta(1-p^2/k^2)$ turns out to be technically very
convenient, since the momentum integration in Eq.~(\ref{eq:flow1}) can be
performed analytically,
\begin{align} 
	\partial_kW_k(\phi)
	=&-\frac{k}{4\pi}
	\frac{W''_k(\phi)}{k^2+W''_k(\phi)^2}.
	\label{eq:flow3}
\end{align}
In order to calculate the bosonic potential $V(\phi)=\frac12 W'(\phi)^2$, we
only need the derivative of the superpotential. The corresponding flow is
\begin{align}
		\partial_kW'_k(\phi)
	=&-W_k'''(\phi)\frac{k}{4\pi}
	\frac{k^2-W_k''(\phi)^2}{(k^2+W_k''(\phi)^2)^2}.\label{eq:flow3a}
\end{align}
This equation exhibits a particularity for any finite value of $k$, as the
sign of the flow depends on whether $W_k''(\phi)^2$ is smaller or larger than $k^2$.
For large $\phi$, we
generally expect both $(W_k'')^2$ and $W_k'''$ to be large and positive for
$\Z_2$ symmetric systems. In this case, the flow for large $\phi$ tends to
deplete the height of the potential. Of course, for $\phi\to\infty$, we expect
the denominator of Eq.~(\ref{eq:flow3a}) to win out over the numerator, such
that the flow vanishes at large field amplitudes. For small $\phi$, or, more
generally, in the vicinity of local or global minima of $W_k'$, there can be
an inner domain where $(W_k'')^2<k^2$. For convex potentials $W_k'$ with
$W_k'''>0$, the flow is negative here, resulting in the tendency to flatten
out this inner part of the potential $W_k'$. As the curvature $W_k'''$ is
related to the masses of the excitations, the flow shows a clear tendency to
small masses if an inner domain with $(W_k'')^2<k^2$ exists. As it will turn
out later, this is a characteristic property of the supersymmetry-broken
phase.

Let us note in passing that the regulator used in the present section can lead
to artificial divergences at higher orders in the derivative expansion, e.g.,
when a wave function renormalization is included.  Then a stronger regulator
in the IR is needed; see, App.~\ref{sec:strongReg1} for calculations at
next-to-leading order in the derivative expansion.

%======================================================
\section{Fixed-point structure}
%======================================================
\label{sec:FixedPointStructure}

In this section, we investigate the fixed-point structure of the RG flow. We
first concentrate on the local-potential approximation and later include
next-to-leading-order terms in the derivative expansion. In fact, it turns out
that there is a qualitative difference of the fixed-point superpotentials
between the different orders. Similar observations are known from
two-dimensional bosonic theories \cite{Morris:1994jc,Neves:1998tg} and are a
particularity of two-dimensional systems. Still, the local-potential flow is
interesting in its own right. Its IR flow is also quantitatively relevant for
studies of the phase diagram, see Sect.~\ref{sec:gaussian-fixed-point}.

Since RG fixed-point studies require a scaling form of the flow equation, we
switch to dimensionless quantities $w$ and $t$ defined by
$W_k(\phi)=kw_t(\phi)$ and $t=\ln(k/\Lambda)$. In two dimensions, a scalar
field is dimensionless such that no dimensionful rescaling is required. The
flow equation \eqref{eq:flow3} for the dimensionless quantities reads
\begin{align} 
	\partial_tw_t(\phi)+w_t(\phi)
	=&-\frac{1}{4\pi}
	\frac{w''_t(\phi)}{1+w''_t(\phi)^2}
	\label{eq:flow3b}
\end{align} 
with $\partial_t=k\partial_k$.  The fixed points are characterized by
$\partial_tw'_\ast=0$. In the following, we solve the fixed-point equation
by various methods.
       
%======================================================
\subsection{Polynomial expansion}\label{polexpansion}
%======================================================

For small values of the field, a polynomial approximation for $w'_t$ is
justified. If $w'_t(\phi)$ is an even function at the cutoff scale then it
remains even at all scales. Its expansion reads
$w'_t(\phi)=\lambda_t(\phi^2-a_t^2)+b_ {4,t}\phi^4
+b_{6,t}\phi^6+b_{8,t}\phi^8+\dots$. The dimensionless couplings
$\lambda_t,b_{t,i}$ relate to the bare couplings $\bar\lambda,\bar b_i$ through
$\bar\lambda=k\lambda_t$ and $\bar b_{i}=k b_{t,i}$. $\bar a$ is dimensionless,
therefore we have $\bar a=a_t$.

 Expanding both sides of the flow equation
in terms of $\phi$, a comparison of coefficients leads to the following system
of coupled ordinary differential equations:
\begin{align}
\partial_t a^2_t=&\frac{1}{2\pi}-\frac{6\lambda^2_t\cdot a^2_t}{\pi}
		+a_t^2\frac{3b_{4,t}}{\pi\lambda_t}\nonumber\\
   \partial_t\lambda_t=&-\frac{3 b_{4,t}}{\pi
   }+\frac{6 \lambda_t^3}{\pi
   }-\lambda_t\nonumber\\
   \partial_tb_{4,t}=&-\frac{15b_{6,t}}{2 \pi}+\frac{60 b_{4,t}\cdot
   \lambda_t^2}{ \pi}-\frac{ 40  \lambda_t^5}{ \pi}-b_{4,t}
\label{eq:Pol1}
   \\
   \partial_tb_{6,t}=&-\frac{14
   b_{8,t}}{\pi   }-\frac{560
    b_{4,t} \lambda_t^4}{\pi }+\frac{168 b_{4,t}^2\lambda_t}{\pi
   }\nonumber\\&+\frac{126 b_{6,t} \lambda_t^2}{\pi
   }+\frac{224\lambda_t^7}{\pi
   }-b_{6,t}\nonumber\\
   &\vdots\nonumber\\
\pa_t b_{2n,t}=&-\frac{(n+1)(n+2)}{4\pi}\,b_{n+2,t}+f_{2n}\left(\lam_t,b_{4,t},
\ldots,b_{2n,t}\right).\nonumber
\end{align}
Note that only $w''$ enters the right-hand side of the flow equation
\eqref{eq:flow3b}; in particular, the lowest coupling constant $a^2_t$ does
not enter the equations for the higher order couplings.

\begin{figure}
\includegraphics[width=.9\columnwidth]{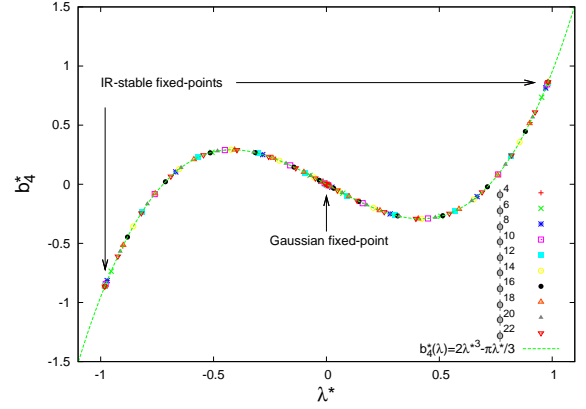}
\caption{Projection of the coefficients of all fixed points for different
truncations on
the plane of the couplings $\lambda$ and $b_4$.\label{fig:lineFixedPoints}}
\end{figure}

At a fixed point, the coupling constants, marked by an asterisk, become scale
invariant such that the left-hand sides in Eq.~\eqref{eq:Pol1} vanish. The
corresponding system of equations
\begin{equation}
 b^*_{2n+2}=\frac{4\pi}{(n+1)(n+2)}\,f_{2n}\left(\lam^*,b^*_4,
\dots,b^*_{2n}\right)\label{eq:Pol5}
\end{equation}
can be solved iteratively due to the triangular form of the system of flow
equations \eqref{eq:Pol1}.  At a fixed point, $b^*_{2n}$ is a polynomial of
order $2n+1$ in $\lam^*$.  Because of the $\Z_2$ symmetry of the system of
equations, these polynomials are odd.  In particular,
$b^*_4(\lam^*)=2\lambda^{*3}-\pi\lambda^*/3$ such that
the projection of any fixed-point solution on the subspace
defined by the couplings $\lam^*$ and $b^*_4$ fall onto the
curve $b_4^*(\lam^*)$ depicted in Fig.~\ref{fig:lineFixedPoints}.
Inserting $b_4^*(\lam^*)$ into the first equation in \eqref{eq:Pol1}
yields $(a^*)^2=1/2\pi$. 
Later, we shall see that $a^2_t$ defines an IR unstable
direction near a fixed point.

Let us truncate the polynomial expansion of the flow equation for $w_t$ and
keep only terms up to order $\phi^{2n+1}$ in \eqref{eq:flow3b}. This is
equivalent to keeping the lowest $n$ fixed point equations which yield
$b_{2m}^*(\lam^*)$ for $m\leq n$. In the $n$th equation, the higher-order
coupling $b_{2n+2}$ occurs which we set to zero at the fixed point,
$b^*_{2n+2}=0$.\protect\footnote{This prescription is not unique. Alternatively, we
  could set $b_{2n+2}^\ast$, for instance, equal to its perturbative one-loop
  value. In any case, the choice $b_{2n+2}^\ast=0$ used here is
  self-consistent in the sense that the equations of $b_{\leq 2n}$ are closed
  and do not depend on further input.} This leads to the polynomial equation
\begin{equation}
f_{2n}(\lam^*)=
f_{2n}\left(\lam^*,b_4^*(\lam^*),\dots,b^*_{2n}(\lam^*)\right)=0,\label{eq:Pol7}
\end{equation}
where the solutions $b_{2m}^*(\lam^*)$ are to be inserted.  With
\textsc{Mathematica}, we have checked up to order $2n=22$ that all $2n+1$
roots of the odd polynomial $f_{2n}(\lam^*)$ of order $2n+1$ are real.

Due to the underlying $\Z_2$ symmetry, the remaining fixed points of the
truncated system come in pairs $\pm(\lam^*,b_4^*,\dots,b_{2n}^*)$. Hence, we
find $n$ independent nontrivial solutions to the fixed-point equations in
addition to the Gau\ss ian fixed point where all couplings vanish. As
discussed in the next subsection, only one of these solutions belongs to an
infrared stable fixed point with all but one eigenvalues in the stability
matrix being positive (i.e., all but one critical exponents being negative).
It turns out that this infrared-stable fixed point corresponds to the largest
root of Eq.~(\ref{eq:Pol7}), as indicated in Fig.~\ref{fig:lineFixedPoints}.
With increasing order of the polynomial truncation the root belonging to the
IR-stable fixed point converges to $\lam_{\rm crit}\simeq 0.9816$. Roots
belonging to any other fixed point are bounded by
\begin{equation}
-\lam_{\rm crit}<\lam^\ast< \lam_{\rm crit}.\label{eq:Pol9}
\end{equation}
%
%%%%%%%%%%%%%%%%%%%%%%%%%
\begin{figure*}
\includegraphics[width=.95\columnwidth]{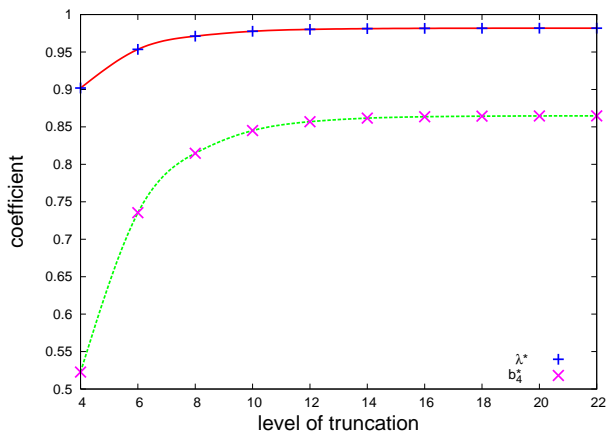}\hfill
\includegraphics[width=.95\columnwidth]{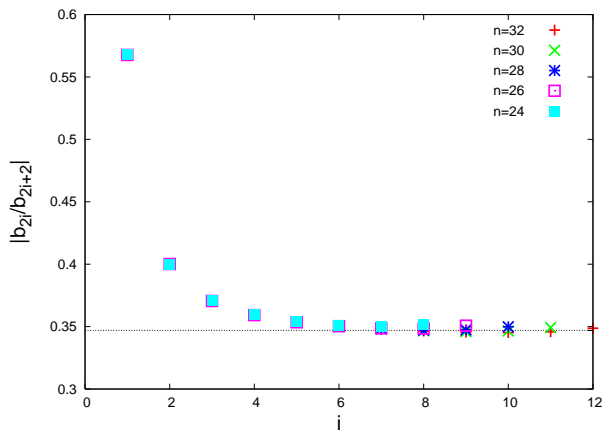}
\caption{Left panel: The first two coefficients $\lambda$ and $b_4$ of the IR
  stable fixed point for different truncations. Right panel: Ratio of
  successive couplings of the IR-stable fixed point yielding an estimate of 
  the radius of convergence.
\label{fig:IRstabilerFixpkt}}
\end{figure*}
In Fig.~\ref{fig:IRstabilerFixpkt} (left panel), we have plotted the values
for the two lowest coefficients $\lam^*$ and $b_4^*$ at the infrared-stable
fixed point in the polynomial expansion for different orders of truncations.
We observe a rapid convergence with increasing order of the polynomial
approximation. This suggests that the polynomial approximation to the
superpotential in the local potential approximation has acceptable convergence
properties.

In Fig.~\ref{fig:IRstabilerFixpkt} (right panel), we have plotted the
inverse ratio of successive couplings at the infrared-stable fixed point for
increasing truncation order. This facilitates an estimate of the radius of
convergence for a series expansion of $w^{\prime}_\ast$ in powers of $\phi^2$,
\[r_{\rm
  con}^2=\lim_{i\to\infty}\Big\vert\frac{b_{2i}^\ast}{b_{2i+2}^\ast}\Big\vert.\]
An extrapolation leads to the approximate value $r^2_{\rm con}\simeq 0.35$
such that for $\vert\phi\vert<r_{\rm con}$ the polynomial
$\lam^*(\phi^2-a^2)+\sum_2^n b^*_{2m}\phi^{2m}$ converges to the IR-stable
fixed point solution $w^{\prime}_\ast$.  In Table \ref{tab:ExponIRstabilA},
the coefficients in the polynomial approximations to $w^{\prime}_\ast$ at the
IR-stable fixed point are listed. Note that the fixed-point values are
generically regulator dependent, and thus are not directly related to physical
quantities.

\begin{table}
\begin{center}
\begin{footnotesize}
\begin{ruledtabular}
 \begin{tabular}{cccccccc}
 & & \multicolumn{6}{c}{Coefficients at IR-fixed point}\\\cmidrule(rl){2-8}
 $2n$&$\lambar^\ast$&$\bbar_4^\ast$&$\bbar_6^\ast$&$\bbar_8^\ast$
 &$\bbar_{10}^\ast$&$\bbar_{12}^\ast$&$\bbar_{14}^\ast$
 	 \\\hline 
2&0.7236\\
4&0.9019& 0.5227\\
6&0.9535& 0.7354& 0.8372\\
8&0.9711& 0.8148& 1.199& 1.694\\
10&0.9777& 0.8451& 1.345&2.420& 3.801\\
12&0.9802& 0.8570& 1.402& 2.716& 5.401& 9.030\\
14& 0.9812& 0.8617& 1.425&2.836& 6.054& 12.77&22.23
\end{tabular}
\end{ruledtabular}
\end{footnotesize}
\caption{The coefficients of the IR-stable fixed point potential for different truncations.
\label{tab:ExponIRstabilA}}
\end{center}
\end{table}

%======================================================
\subsubsection{Stability analysis and critical exponents}
%======================================================

\begin{table*}
\begin{center}
\begin{ruledtabular}
\begin{scriptsize}
\begin{tabular}{c|cccccccc}
 $\lambda^\ast$&\multicolumn{8}{c}{ Critical exponents $\theta^I$
 }\\\hline
 $\pm.9816$&$-1.54$& $-7.43$&$- 18.3$&$ -37.3$&$- 68.9$&$- 120$&$- 204$&$-351$\\
 $\pm.8813$&$ 6.16$&$-1.64$&$-9.82$&$-25.6$&$-52.5$&$-96.9$&$-170$&$-300$\\
 $\pm.7131$&$21.4$&$4.37$&$-1.57$& $-11.1$&$ -30.1$&$ -63.3$& $-120$&$-223$\\ 
 $\pm.5152$&$ 28.7$&$ 13.3$&$ 3.33$&$ -1.39$&$ -11.6$&$ -32.8$&$ -71.7$&$-145$\\ 	
 $\pm .3158$& $20.0-4.55$\;i&$20.0+4.55$\;i&$ 8.40$
   & $2.57$&$-1.14$ &$-11.6$ & $-34.3$ & $-80.4$
 	\\
$\pm.1437$& $11.2+9.02$\;i
& $11.2-9.02$\;i&
  $8.63$&
   $5.19$&
   $1.95$&
  $-.842$& $-11.1$&	$-35.7$  
\\
$\pm.0322$& $4.20+1.18$\;i
& $4.20-1.18$\;i& $2.86$&$ 2.72+6.47$\;i
& $2.72-6.47$\;i
&$ 1.47$&$ -.540$&$- 10.5$
 \\
$\pm .0003$&$ 1.57+.125$\;i
&$1.57-.125$\;i&$1.43+.702$\,i
&$1.43+.702$\;i

&$ 1.14$&$.542+.982$\;i
&$ .542+.982$\;i
&$ -0.221$
 \\
 0&1& 1&1& 1& 1& 1& 1& 1
\end{tabular}
\end{scriptsize}
\end{ruledtabular}
\caption{Critical exponents $\theta^I$ (negative eigenvalues of the stability
  matrix) for a polynomial truncation at $2n=16$ for the nine different fix
  points in the local-potential approximation.
 The first exponent $\theta^0=1$ which is common to all fixed points is not
 shown here.
\label{tab:EigsStabMatrixN16}}
\end{center}
\end{table*}

Whereas the values of the fixed-point couplings are regulator dependent, the
{\em critical exponents} are universal and give rise to a classification of
the fixed points. The critical exponents are defined as the negative
eigenvalues $\theta^I$ of the stability matrix at the fixed point, 
\begin{equation}
B_i{}^j= \frac{\partial (\partial_t b_i)}{\partial b_j} \bigg|_{b=b^\ast},
\quad B_i{}^j v_j^I = -\theta^I v_i^I,
\label{eq:1}
\end{equation}
where we have set $b_0=a_t^2$, $b_2=\lambda$, and $I$ labels the different
critical exponents $\theta^I$ and eigendirections $v_i^I$. Critical exponents
with positive real part correspond to RG relevant directions, whereas
exponents with negative real part mark irrelevant directions. Inserting the
flow of $\lambda_t$ into that of $a_t^2$, cf. Eq.~(\ref{eq:Pol1}), we obtain
\begin{equation}
\partial_t a_t^2 = \frac{1}{2\pi} -a_t^2 - \frac{a_t^2}{\lambda_t} \,
\partial_t \lambda_t. \label{eq:1b}
\end{equation}
We observe that the 00-component of the stability matrix at any fixed-point
yields, $B_0{}^0=-1$. This together with the fact that the remainder of the
first column vanishes, $B_{i\geq 1}{}^{0}=0$, implies that $a_t^2$ is always an
eigendirection of $B_i{}^j$ with corresponding critical exponent
$\theta^0=1$. Note that this result is manifestly regulator independent and
thus universal. We conclude that any fixed point of the superpotential in the
local-potential approximation has at least one RG relevant direction. In
analogy with potential flows near the Wilson-Fisher fixed point of Ising-like
systems, we introduce the following notion for the leading critical exponent
corresponding to this relevant direction:
\begin{equation}
\nu_W=\frac{1}{\theta^0}=1.
\end{equation}
Even though $\nu_W$ plays the same role for the superpotential flow
as the critical exponent $\nu$ does for the potential flow in Ising-like
systems, it should be stressed that $\nu_W$ does not correspond to the scaling
exponent of the correlation length (as $\nu$ does in Ising-like systems). We
will later see that $\nu_W$ quantifies certain properties of the phase
diagram.
 
Depending on whether we study the UV or IR flow of the system, the fixed
points have a different meaning. Towards the UV, any of the fixed points which
we have found can be used to define a UV completion of the model in the sense
of Weinberg's asymptotic safety scenario \cite{Weinberg:1976xy}. All relevant
directions emanating from the fixed point span the critical hypersurface. The
dimensionality of this critical surface, i.e., the number of critical exponents with
$\text{Re}\,\theta^I\geq 0$, corresponds to the number of physical parameters which
have to be fixed in order to unambiguously define the flow towards the
IR.\protect\footnote{For marginal directions with $\text{Re}\,\theta^I=0$, the flow in
  the fixed-point regime has to be studied beyond linear order. Depending on
  the sign of the first non-vanishing order, these directions are again either
  marginally relevant or irrelevant.} Once these initial conditions to the
flow are provided, any other quantity or correlation function can be predicted
within the theory. Since $\theta^0=1$, we conclude that any UV completion has
at least one physical parameter.

Within the local-potential approximation at order $2n=16$ of the polynomial
expansion, the critical exponents $\theta^{I\geq 1}$ are given in Table
\ref{tab:EigsStabMatrixN16} for all 17 fixed-point potentials. By using a
different regulator, we check in App.~\ref{app:reg} that the regulator
dependence of the relevant positive critical exponents is rather small (up to
10\%\ or much less), which confirms the reliability of the present
truncation. Classifying the fixed-point potentials by the slope $\lambda^\ast$
of the potential $w_t'$ as a function of $\phi^2$ at $\phi^2=0$, the number of
relevant directions increases as the slope $\abs{\lambda^\ast}$ decreases. The
different fixed-point potentials in the local-potential approximation -- if
they persist to higher truncation orders -- thus correspond to different UV
completions of the present system with increasing physical parameters. These
different UV completions thus define different nonperturbatively renormalized
Wess-Zumino models in two-dimensions.

As for the flow towards the IR, the fixed points can generically be related
with critical points in the phase diagram of the system. Since the relevant
directions are IR repulsive, fine-tuning the relevant direction to the fixed
point corresponds to tuning the system onto its critical point. In this sense,
the relevant direction corresponding to $a_t^2$ with $\nu_W=1/\theta^0=1$ is similar
to the temperature parameter in Ising-like systems (or a mass parameter in
O($N$)-type relativistic models). For instance, in the domain of attraction of
the maximally IR-stable fixed point with only $a_t^2$ as relevant direction,
the tuning of $a_t^2$ distinguishes between the supersymmetric and
symmetry-broken phases of the model. More generally, if a system is in the
domain of attraction of a fixed-point with $N$ relevant directions, the phases
of broken and unbroken supersymmetry are separated by an $N$-dimensional
hypersurface in the space of couplings.

There is one important difference to Ising-like systems: the coupling $a_t^2$
associated with the one common relevant direction does not feed back into the
flow of the higher-order couplings. Therefore, the remaining couplings are
attracted towards the maximally IR-stable fixed point for any regular
trajectory irrespective of the flow of $a_t^2$.\protect\footnote{Irregular
  trajectories can run to infinity at a finite value of $k$. These divergences
  are either physically meaningless or signal the breakdown of the
  truncation.} We conclude that the maximally IR-stable fixed point governs
the flow towards the IR of $w_t''(\phi)$ in the domain, where the polynomial
expansion is valid. 

Let us finally mention that the critical exponent $\nu_W=1/\theta^0=1$
receives corrections at higher orders in the derivative expansion,
cf. Eq.~\eqref{eq:superscaling}. Still the relevance of the maximally IR-stable
fixed point for the IR flow of the potential persists.

In Table \ref{tab:EigsStabMatrixIRstable}, we collected the eigenvalues
(negative critical exponents) for the
maximally IR-stable fix point for different truncations.%
\begin{table}
\begin{center}  
\begin{ruledtabular}
\begin{footnotesize}
\begin{tabular}{cc|cccccccc}
$2n$&$\lambda^\ast$&\multicolumn{8}{c}{eigenvalues of the stability matrix}\\\hline
4&$\pm$0.9019&&&&&&&16.35& 1.846\\
6&$\pm$0.9535&&&&&&42.32& 12.00& 1.716\\
8&$\pm$0.9711&&&&&79.83& 30.67& 9.951& 1.635\\
10&$\pm$0.9777&&&&129.1& 58.61& 25.05& 8.794& 1.588\\
12&$\pm$0.9802&&&190.6& 96.49& 47.97& 21.75& 8.101& 1.561\\
14&$\pm$0.9812&&264.5& 144.8& 79.50& 41.51& 19.67& 7.680& 1.546\\
16&$\pm$0.9816&351.2& 204.1& 120.3& 68.90& 37.25& 18.30& 7.427& 1.539\\
\end{tabular}
\end{footnotesize}
\end{ruledtabular}
\caption{Eigenvalues of the stability matrix (negative critical exponents) of
  the maximally IR-stable fix point for different polynomial truncations in
  the local-potential approximation. The first exponent $\theta^0=1$ is not shown
  here. For the first subleading exponent (last column), the polynomial
  expansion shows a satisfactory convergence. \label{tab:EigsStabMatrixIRstable}}
\end{center}
\end{table}

%======================================================
\subsection{Solving of the nonlinear differential equation}
%======================================================

For large values of the scalar field, the polynomial truncation is not valid
anymore. Hence, we consider here the full nonlinear ordinary differential
equation that describes the fixed point potential in the local-potential
approximation. In the polynomial approximation, the potential $w_t$ and
$w_t^\prime$ contain the IR-unstable coupling $a_t$ which does not flow into
the fixed point. Upon differentiation of the fixed-point equation for
$w_\ast'(\phi)$,
\begin{align}
	w_\ast '=-\frac{w_\ast '''}{4\pi}
\frac{1-w_\ast''^{\,2}}{(1+w_\ast ''^{\,2})^2},
	\label{eq:fixedpoint1}
\end{align}
with respect to $\phi$, we arrive at the following fixed-point equation for
the $a_*\,$-independent function $w_*''(\phi)\equiv u(\phi)$,
\begin{equation}
(1-u^4)u''=2 u^{\prime\,2}\,(3-u^2)\,u- (1+u^2)^3\,4\pi u.
\label{eq:fixedpoint5}
\end{equation}
Since $a_t^2$ does not appear in Eq.~\eqref{eq:fixedpoint5}, we expect to find
a fully IR-stable solution to this nonlinear differential equation (in
addition to further solutions with IR unstable directions).

As before, we consider odd solutions $u(\phi)$ which are fixed by the initial
conditions $u(0)=0$ and a finite value for $u'(0)$.
In fact, we find a continuum of oscillatory solutions that are defined for all
values of $\phi$. In addition, we identify solutions that hit the singular
line $u(\phi)=1$ of the differential equation (\ref{eq:fixedpoint5}) but can
be continued without cusps.  Solutions of the second class exist only in a
finite $\phi$ range.

%======================================================
\subsubsection{Oscillating solutions}
%======================================================

  \begin{figure}
\includegraphics[width=.9\columnwidth]{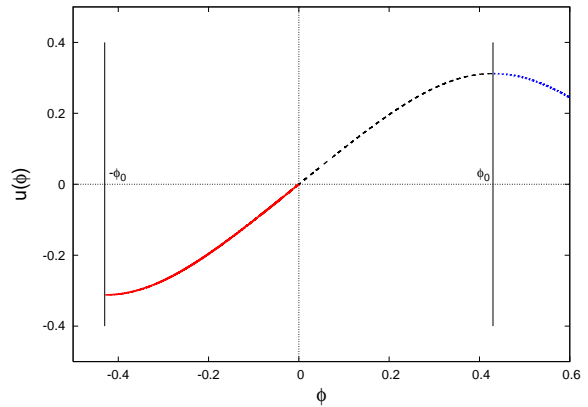}
\caption{Illustration of the discussion of oscillating
  solutions, see text. \label{fig:oszillatingSol}}
\end{figure}

Let us integrate the fixed-point equation \eqref{eq:fixedpoint5} with initial
conditions at $-\phi_0<0$ (see Fig.~\ref{fig:oszillatingSol}):
\begin{equation*}
u(-\phi_0)=u_0,\quad u'(-\phi_0)=0\mtxt{where}
 -1<u_0<0.
\end{equation*}

The initial point is an extremum of the solution.
The differential equation~\eqref{eq:fixedpoint5} 
implies that this extremum is indeed a minimum 
such that $u$ approaches the $\phi$ axis away
from $-\phi_0$. Actually it must intersect the axis 
at some point. Otherwise, it would possess a 
maximum at some $\phi_1$ with $u(-\phi_0)<u(\phi_1)<0$ 
or it would monotonically approach the $\phi$ axis without crossing it.  But a
maximum with $-1<u(\phi_1)<0$ contradicts the differential equation
\eqref{eq:fixedpoint5} which implies that $u''(\phi_1)>0$. Also, if $u$
approached the $\phi$ axis monotonically from below then near the axis the
differential equation would imply $u''\approx -4\pi u>0$.  Therefore, the
slope of $u$ would increase with increasing $\phi$ such that finally $u$ would
intersect the $\phi$ axis.

We conclude that $u$ must intersect the $\phi$ axis and we may
assume that this happens at $\phi=0$, such that the solution 
starts off at the minimum at $-\phi_0<0$ and hits the axis at the origin. 
This solution extends to an odd solution owing to the symmetry 
$u(\phi)\to -u(-\phi)$  of the differential equation and hence
has a maximum at $\phi_0>0$ with $0<u(\phi_0)=-u(-\phi_0)<1$. 
Because of the translational invariance $\phi\to\phi+c$ and the symmetry 
$u(\phi)\to u(-\phi)$ of the fixed point equation
the solution must be symmetric relative to $\phi_0$, i.e. $u(\phi_0-\phi)=
u(\phi_0+\phi)$. This proves that every solution with
$-1<u(\hbox{extrema})<1$ is periodic and takes its values between 
$-1$ and $1$. The lines $u(\phi)=\pm 1$ repel solutions oscillating
in the strip $-1<u<1$ as long as the slope $u'(0)=\gamma$ is less than
$\gamma_{\rm crit}\simeq1.964$.  Solutions with $\gamma\geq \gamma_{\rm crit}$
hit the singularity at $u=1$.

In a similar fashion, one argues that there exists a second class of
solutions of the differential equation having just one minimum with
$u>1$ or having just one maximum with $u<-1$. Solution in this
class are not periodic and never hit the singular line $u=\pm 1$. Since
they cannot be continuous and antisymmetric they are discarded. 

%======================================================
\subsubsection{Comparison with the polynomial expansion}
%======================================================

\begin{figure*}
	\includegraphics[width=.9\columnwidth]{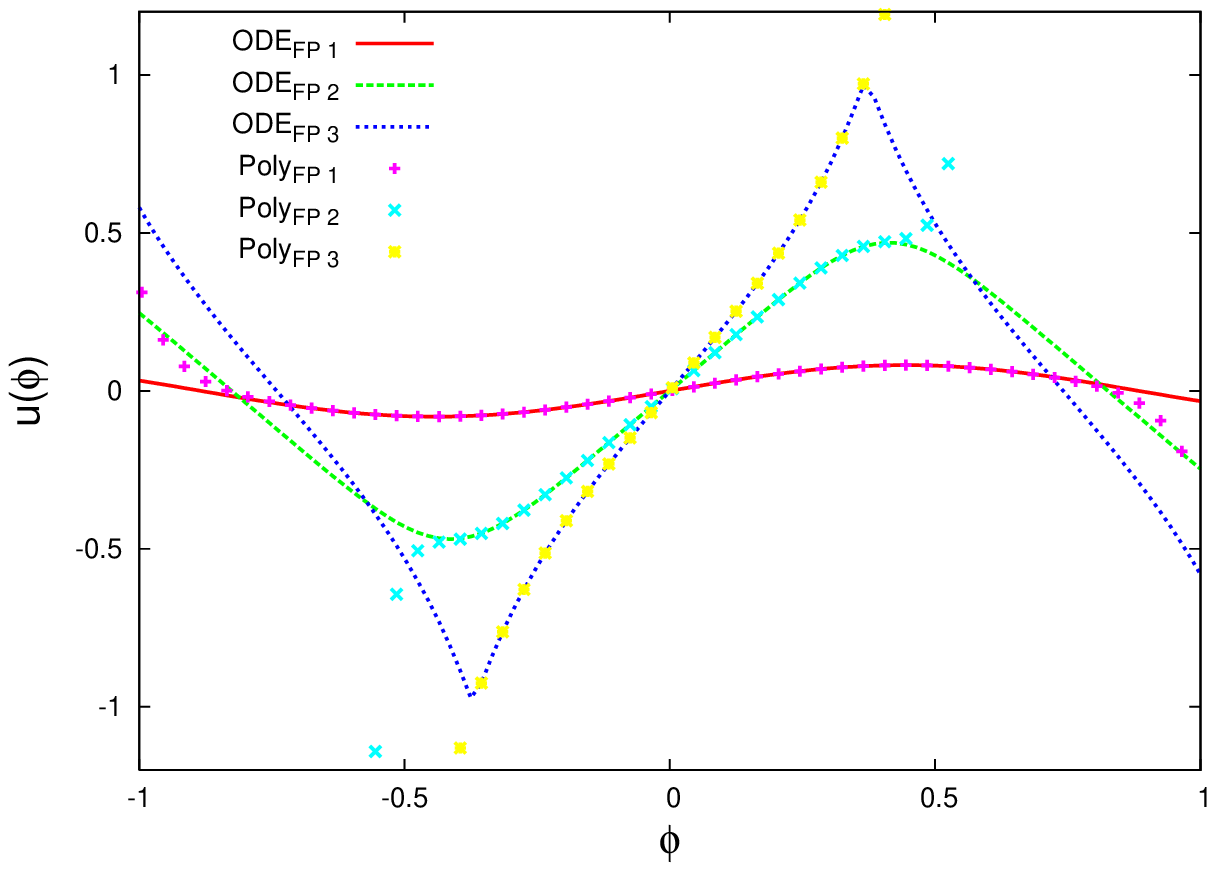}\hfill
	\includegraphics[width=.9\columnwidth]{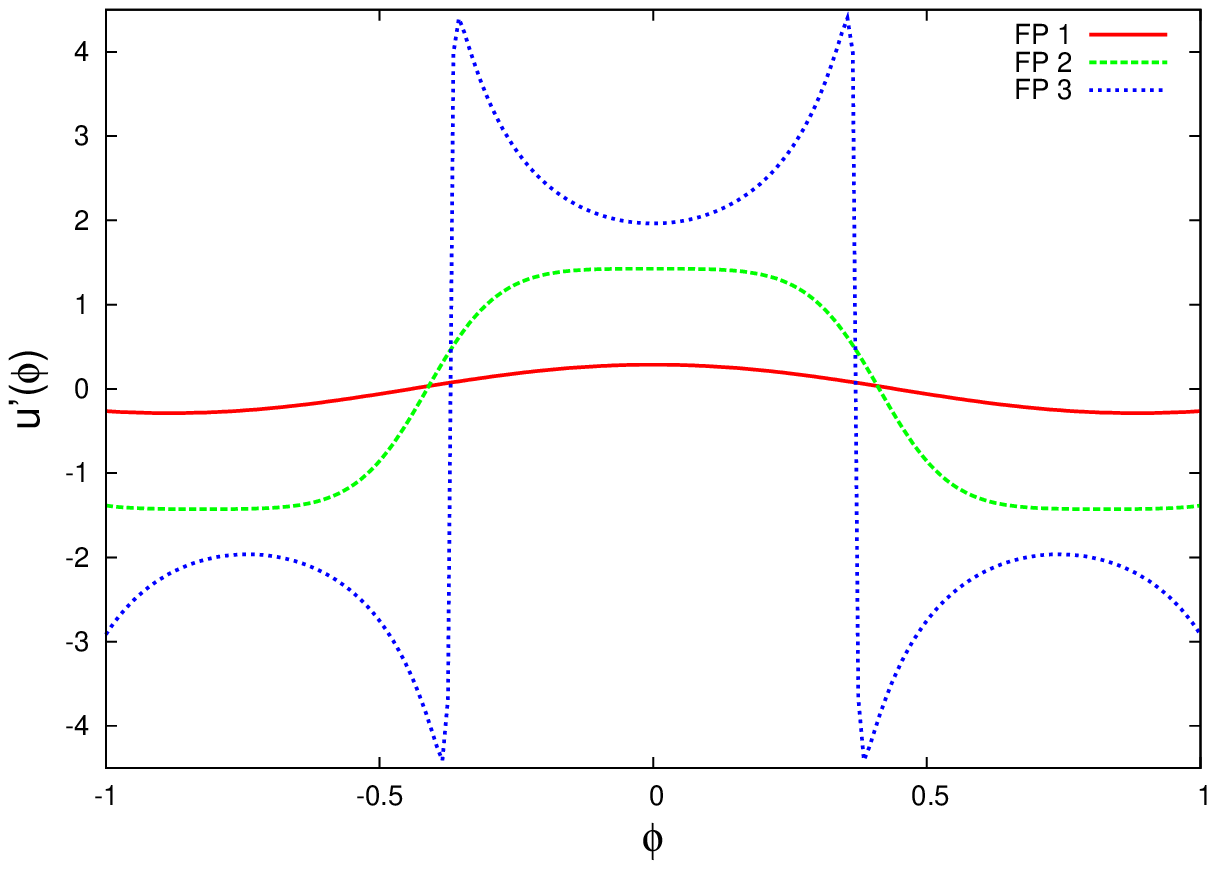}
	\caption{\emph{Left panel:} Comparison between the numerical solution
          to the differential equation (ODE) and the polynomial approximation
          (Poly) to 16th order for three different fixed points. The fixed
          points FP1, FP2, and FP3 have the initial slope $\gamma=0.287,1.4262$
          and $1.963$. The fixed point FP3 is the maximally
          IR-stable fixed-point. \emph{Right panel:} The first derivative of
          the potentials in Fig.~\ref{fig:FitFixPoint}. \label{fig:FitFixPoint}}
\end{figure*}

In the preceding subsection, we have seen in the polynomial expansions of the
truncated system that the slope $\gamma=2\lambda$ is bounded by $2\lam_{\rm
  crit}\simeq 1.964$. This is just the critical value $\gamma_{\rm crit}$ for the
existence of oscillating solutions of the fixed point equation
\eqref{eq:fixedpoint5}.  We conjecture that a polynomial solution belonging to
a fixed point with two or more relevant directions of the truncated systems,
corresponding to a non-maximal root of $f_{2n}$ in \eqref{eq:Pol7}, converges
to the Taylor series of an oscillating solution.  In Fig.~\ref{fig:FitFixPoint} 
(left panel), we have plotted three full numerical
solutions and the corresponding polynomial approximation truncated at
$\phi^{16}$ with the same initial value $\gamma=2\lambda$. For the first half
period, we find an excellent agreement between polynomial approximation and
numerical solution.

%======================================================
\subsubsection{Solution with $u(\phi)=1$ for some field value $\phi$}
%======================================================

Regular periodic solutions only exist for $-\gamma_{\rm crit}<\gamma< \gamma_{\rm
  crit}$.  Increasing the slope at the origin gradually from $0$ to $\gamma_{\rm
  crit}$, the value $u(\phi_{\rm max})$ at the maximum approaches the
singular line $u=1$ and finally hits the singularity at the critical field
$\phi_{\rm crit}\simeq 0.3704$. At the same time the curvature at the maximum tends
to $-\infty$. With increasing order the IR stable fixed points for the
polynomial truncations converge to the Taylor expansion of the solution with
the critical slope $\gamma_{\rm crit}$.

\begin{figure}
	\includegraphics[width=.9\columnwidth]{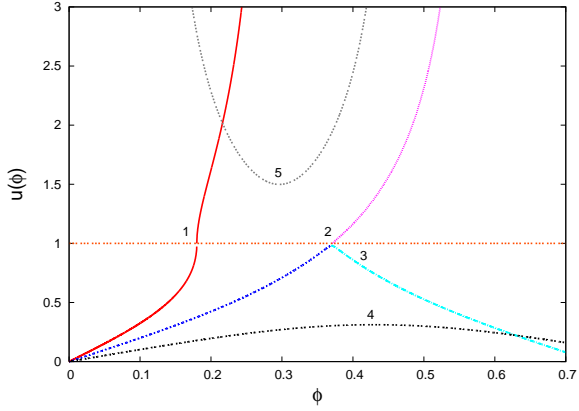}
	\caption{All types of possible solutions to the fixed-point
          differential equation in local-potential approximation: (1)
          $\gamma>\gamma_{\rm crit}$, (2) $\gamma=\gamma_{\rm crit}$, finite
          $\phi$ range, (3) $\gamma=\gamma_{\rm crit}$, oscillating solution,
          (4) $\gamma<\gamma_{\rm crit}$, oscillating
          solution, (5) solution with just one extremum
	 \label{fig:TypesOfSolutions}}
\end{figure}

In order to study the solutions near the singular line,
we insert the Taylor expansion $u(\phi_{\rm crit}+\delta\phi)=u(\phi_{\rm crit})+ a_1 \delta\phi+a_2 \delta\phi^2/2\,+\dots$ with 
$u(\phi_{\rm crit})=1$ and compare coefficients. One sees that the expansion 
coefficients are finite if $a_1=u'(\phi_{\rm crit})=\pm\sqrt{8\pi}$. If
this condition is not met, the solution hits the singular line $u(\phi)=1$
with infinite slope, as can be seen by studying the differential equation for
the inverse function $\phi(u)$.  The behavior of a solution depends in an
essential way on $\gamma$: if the initial slope is less than $\gamma_{\rm crit}$
then the solution is smooth and periodic, if the initial slope is $\gamma_{\rm
  crit}$ then it hits the singular line $u(\phi)=1$ with slope $\sqrt{8\pi}$
and if the initial slope is bigger than $\gamma_{\rm crit}$ then the solution hits
the singular line vertically; see Fig.~\ref{fig:TypesOfSolutions}.  If we
viewed the solutions hitting the singular line as parametric continuation of
the periodic solutions as $\gamma$ approaches $\gamma_{\rm crit}$, we would
reflect the solution at the singular line, similarly to the solution FP3 in
Fig.~\ref{fig:FitFixPoint}, left panel.  For the maximally IR-stable fixed-point
solution with initial slope $\gamma_{\rm crit}$ the slope at $\phi_{\rm crit}$
would then jump from $\sqrt{8\pi}$ to $-\sqrt{8\pi}$, as shown in
Fig.~\ref{fig:FitFixPoint}, right panel, where we depicted the function
$u'(\phi)=w'''(\phi)$ for three different values of the initial slope
$\gamma=2\lambda$.

However, there is a way to extend the solutions hitting the singular line
without cusps. To see this more clearly we note that $v=1/u$ fulfills almost
the identical fixed-point equation as $u$,
\begin{equation}
(1-v^4)v''=2 v^{\prime\,2}\,(3-v^2)\,v- 4\pi (1+v^2)^3/v.
\label{eq:fixedpoint7}  
\end{equation}
Upon approaching the singular line $u\to 1$ and $v\to 1$, the
two equations \eqref{eq:fixedpoint5} and \eqref{eq:fixedpoint7} become
identical. This implies that near the singular line the reflection
at the singular line maps solutions into solutions. Thus, all solutions
with $\gamma\geq \gamma_{\rm crit}$ hitting the singular line vertically or with
slope $\sqrt{8\pi}$ can be extended without cusps beyond the singular 
line. 

We have studied these solutions for large values $u\gg 1$.  It is not
difficult to see that there exist no solutions with $u\sim \phi^\alpha$ for
large $\phi$. (This will become different at next-to-leading order in the
derivative expansion.) We find the asymptotic solution
\begin{align}
	u_{\rm as}(\phi)=e^{\text{erf}^{-1}\left(\pm2 \sqrt{2}\; e^{-\frac{c_1}{8 \pi }} 
		\left(\phi+c_2\right)\right){}^2-\frac{c_1}{8 \pi }}
\end{align}
with $c_1\simeq-20.02$ and $c_2\simeq-0.423$ for the maximally IR-stable  fixed
point  extended without cusp beyond the singular line. 
It is finite only for
\begin{equation}
-\phi_{\rm crit,2}<\phi <\phi_{\rm crit,2},\quad
\phi_{\rm crit,2}=0.5823\label{eq:fixedpoint9}.
\end{equation}
An unbounded and cusp-free fixed point solution belongs to
a field  theory with compact target space.
%=====================================================

%======================================================
Let us summarize our findings. 
As a regular oscillating or
a non-differentiable bouncing solution $u(\phi)$ is finite for all values
of the field, this implies a vanishing \emph{dimensionful} 
$W_k''(\phi)=k u(\phi)$ as the scale $k$ is lowered to the 
infrared.
The solutions which penetrate the singular line $u=1$
without a cusp are unbounded from below and above and
confine the field to a finite interval.  
The different types of solutions are depicted in Fig.~\ref{fig:TypesOfSolutions}.

%======================================================
\subsection{Fixed points at next-to-leading order in the derivative expansion}
%======================================================

\subsubsection{Next-to-leading-order flows}

Two dimensional scalar field theories and their supersymmetric extensions
exhibit infinitely many scale invariant fixed-points characterized by the
central charge $c$ of a conformal field theory \cite{Zamolodchikov86}.  In the
local potential approximation to scalar field theories, only periodic,
sine-Gordon type fixed-point solutions are accessible. At next-to-leading
order in the derivative expansion, one finds additional non-periodic
fixed-point solutions \cite{Morris:1994jc,Neves:1998tg}.  Here we find
analogous results for the two-dimen\-sional Wess-Zumino model.

In this subsection, we consider fixed point solutions with scale-dependent but
field-independent wave function renormalization $Z_k$ in the next-to-leading
order approximation. The corresponding flow equations are derived in appendix
\ref{sec:flowWithWaveFkt}. The full next-to-leading-order approximation would
include a field-dependent wave function renormalization $Z_k(\Phi)$.  For
supersymmetric quantum mechanics, this order has been computed in
\cite{Synatschke:2008pv}. Here, we confine ourselves to the simpler
approximation $Z_k(\Phi) \to Z_k(0)\equiv Z_k$.

Let us introduce renormalized fields $\chi$, by rescaling $\phi$ with the wave
function renormalization $\phi\to\chi=Z_k\phi$. This implies a dimensionless
renormalized superpotential $\w_t(\chi)=W_k(\chi/Z_k)/k$. The calculation of
the flow is outlined in App.~\ref{sec:flowWithWaveFkt}; in order to avoid
artificial IR singularities, different regulator shape functions were used in
comparison with the local-potential approximation: $r_1=0$ and
$r_2=(k^2/p^2-1) \theta(1-p^2/k^2)$. For the rescaled quantities, the flow
equations read
\begin{align}
 \label{eq:NextToLead1}	\partial_t\w'_t&+\w'_t-\frac{\eta}{2}(\chi\w'_t)'
	  =		\frac{1}{4\pi}\frac{\w_t'''}{\w_t''^2}\times\\&\left[
			\ln\left({1 +\w_t''^2}\right)\left(1-\frac{\eta}{2}\frac{3  +\w_t''^2}{\w_t''^2}\right)-
			\frac{2\w_t''^2}{1+\w_t''^2}+\frac{3\eta}{2}
			\right],\nonumber
 				  \\
 \label{eq:NextToLead3}\eta:=&-\partial_t \ln
Z_k^2=\frac{1}{4\pi}\left(\frac{\w'''_k}{{\w''_k}^2}\right)^2\times\\&
\left[\frac{\eta\, {\w''_k}^2}{{1+\w_k''}^2} -\eta\ln
\left(1+{\w_k''}^2\right) +\frac{2
  {\w''_k}^4}{(1+{\w_k''}^2)^2}\right]_{\chi=0},\nonumber
\end{align}
where we have dropped the arguments in $\w_t(\chi)$ for simplicity. Since the
anomalous dimension is assumed to be constant in this approximation, we have
projected Eq.~\eqref{eq:NextToLead3} onto $\chi=0$.  Note that the limit
$\w''_t(0)\to 0$ of the right-hand side of Eq.~\eqref{eq:NextToLead3} exists,
yielding
\begin{align}
\eta  &=\frac{4\lambda^2}{\lambda^2+2\pi}.\label{eq:NextToLead4}
\end{align}

\subsubsection{Polynomial expansion and superscaling relation}

The polynomial expansion of the next-to-leading-order superpotential flow
equation in terms of dimensionless renormalized couplings is given in
App.~\ref{sec:flowWithWaveFkt}. For instance, the flow of the renormalized
parameter $a_t^2$ can be written as (cf. Eq.~\eqref{eq:polyeta})
\begin{equation}
\partial_t a_t^2 = \frac{1}{2\pi} \left( 1- \frac{\eta}{4} \right) - \left( 1-
\frac{\eta}{2} \right) a_t^2 - \frac{a_t^2}{\lambda_t}\, \partial_t \lambda_t,
\end{equation}
which is the next-to-leading-order analogue of Eq.~\eqref{eq:1b}. The
renormalized couplings are related to their unrenormalized analogues by
\begin{equation}
\lambda_t=\frac{1}{k} \, \frac{1}{Z_k^3}\, \bar{\lambda}_k, \quad a_t^2 =
Z_k^2 \bar{a}_k^2.
\end{equation}
Similarly to the local-potential approximation, we observe that the
00-component of the stability matrix at any fixed point yields
$B_0{}^0=-(1-\frac{\eta}{2})$, where $\eta=\eta^\ast$ has to be evaluated at
the corresponding fixed point. Since the remainder of the first column
vanishes, $B_{i\geq1}^0=0$, the coupling $a_t^2$ remains always an
eigendirection of $B_i{}^j$ at any fixed point with a critical exponent
$\theta^0=-(1-\frac{\eta}{2})$, implying a {\em superscaling relation}
\begin{equation}
\nu_W\equiv \frac{1}{\theta^0}=\frac{2}{2-\eta}, \label{eq:superscaling}
\end{equation}
where we have again introduced an Ising-like notation for the critical
exponent of the superpotential associated with the $a_t^2$ direction. This is
a remarkable relation as it relates this superpotential exponent with
the anomalous dimension. Recall that in Ising-like systems the thermodynamic
main exponents (i.e., $\alpha$, $\beta$, $\gamma$ and $\delta$) are related
among each other by scaling relations, and can be deduced from the correlation
exponents $\nu$ and $\eta$ by hyperscaling relations. Beyond that there is no
general relation between $\nu$ and $\eta$. The superscaling relation
\eqref{eq:superscaling} thus represents a special feature of the present
supersymmetric model.

We would like to stress that Eq.~\eqref{eq:superscaling} is an exact relation
to next-to-leading order in the supercovariant derivative expansion of the
effective action. In particular, the inclusion of a field-dependent
$Z_k(\phi)$ implying $\eta\to \eta(\phi)$ does not modify the superscaling
relation, since the superscaling relation arises from the expansion in $\phi$
near $\phi=0$. Beyond next-to-leading order, Eq.~\eqref{eq:superscaling}
might, in fact, receive corrections, since higher-derivative operators can
still take influence on the flow of the superpotential mediated by
higher-order interactions between the scalar field and the auxiliary
field. Whether or not these interactions play a role for the superscaling
relation at the fixed points needs to be clarified by future studies.

Within the present next-to-leading-order truncation, a numerical determination
of the critical exponents from the full set of polynomially expanded flow
equations, of course, confirms the superscaling relation to a high
accuracy. Numerical values for $\eta$ and $\nu_W$ at the maximally IR-stable
fixed point for increasing order of polynomial truncations are given in Table
\ref{tab:ExponIRstabil}.%
\begin{table}
\begin{center}
\begin{footnotesize}
\begin{ruledtabular}
\begin{tabular}{cccccccc}
$2n$&2&4&6&8&10&12&14\\\hline
$\eta$    &0.3284&0.4194&0.4358&0.4386&0.4388&0.4387&0.4386\\\
 $1/\nu_W$&0.8358&0.7903&0.7821&0.7807&0.7806&0.78065&0.7807
\end{tabular}
\end{ruledtabular}   
\end{footnotesize}       
\caption{Numerical verification of the superscaling relation
  \eqref{eq:superscaling}: anomalous dimension $\eta$ and the critical
  exponent $1/\nu_W$ of $a^2$ for increasing orders in a polynomial truncation
  evaluated for the maximally IR-stable fixed-point.
\label{tab:ExponIRstabil}}
\end{center} 
\end{table}  
We also observe a rapid convergence of the polynomial expansion, yielding our
best estimates  $\eta\simeq0.4386$ and $1/\nu_W\simeq 0.7807$ for the critical
exponents at the maximally IR-stable fixed point. As the anomalous dimension
is comparatively large, we expect significant quantitative corrections to
arise from higher orders in the derivative expansion. 

As discussed below, the superscaling relation has an immediate physical
consequence for the IR flow of the masses in the supersymmetry broken phase. 

\subsubsection{Fixed points of the nonlinear superpotential flow at next-to-leading order}

In order to go beyond the polynomial expansion, let us first study the
asymptotic behavior of the right-hand side of the flow equation
\eqref{eq:NextToLead1}:
\begin{align}
\w_t''^2\to 0:&\;\frac{\eta-4}{16\pi}\,\w_t'''\quad\hbox{and}\\\quad
\w_t''^2\to\infty:&\;\frac{2-\eta}{8\pi}\frac{\w_t'''}{\w_t''^2}\,\ln(1+\w_t''^2)
.
\end{align}
It turns out to be a self-consistent assumption that these asymptotic
right-hand sides are subdominant in comparison with the left-hand side of
Eq.~(\ref{eq:NextToLead1}) at a fixed point $\partial_t\w'_\ast=0$ both for
small and large values of $\chi$. From this, it follows that $\w'_\ast$ is
proportional to $\chi^{2/\eta-1}$ for large $\w''_\ast$. In particular,
$\w'_\ast$ grows faster than any polynomial for $\eta=0$, in complete agreement
with our previous results in section \ref{sec:FixedPointStructure}. 

\begin{figure*}
	\includegraphics[width=.9\columnwidth]{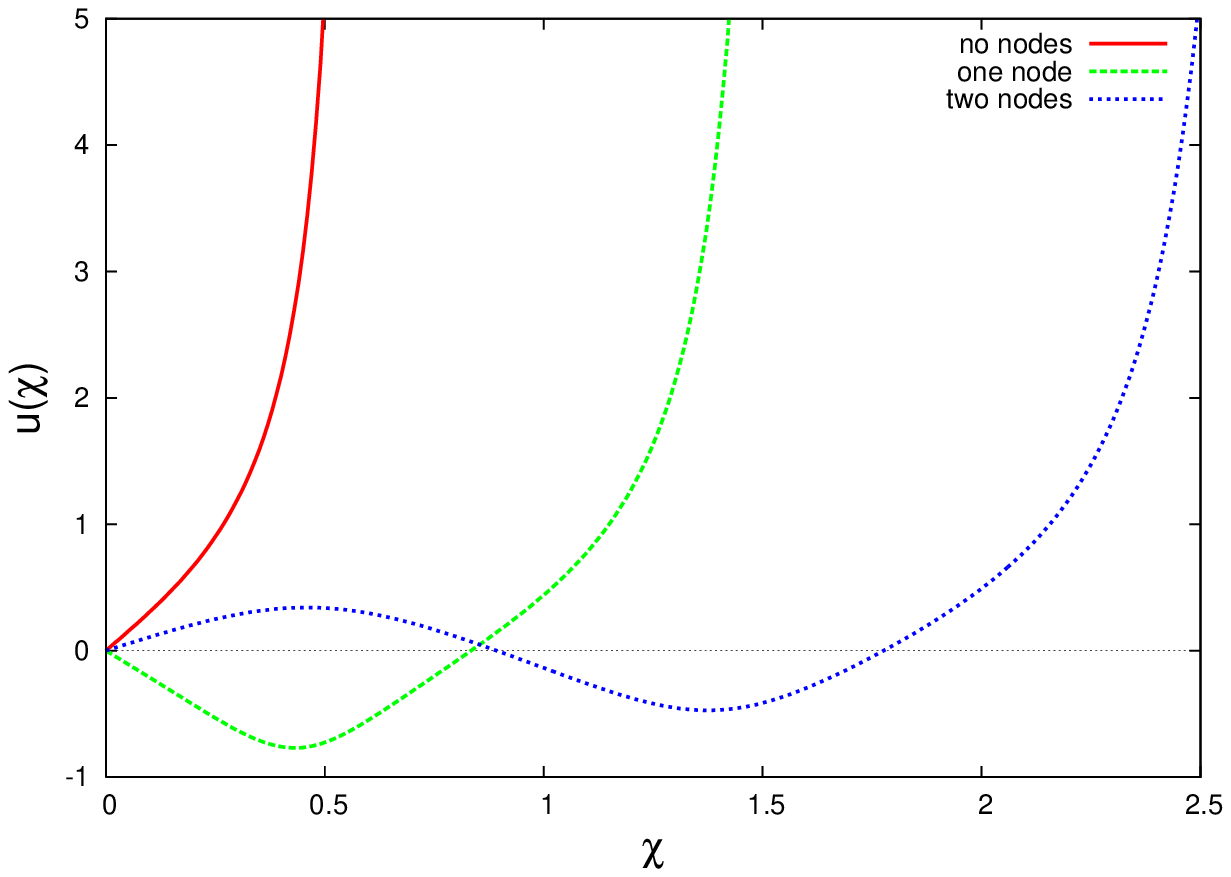}
\includegraphics[width=.9\columnwidth]{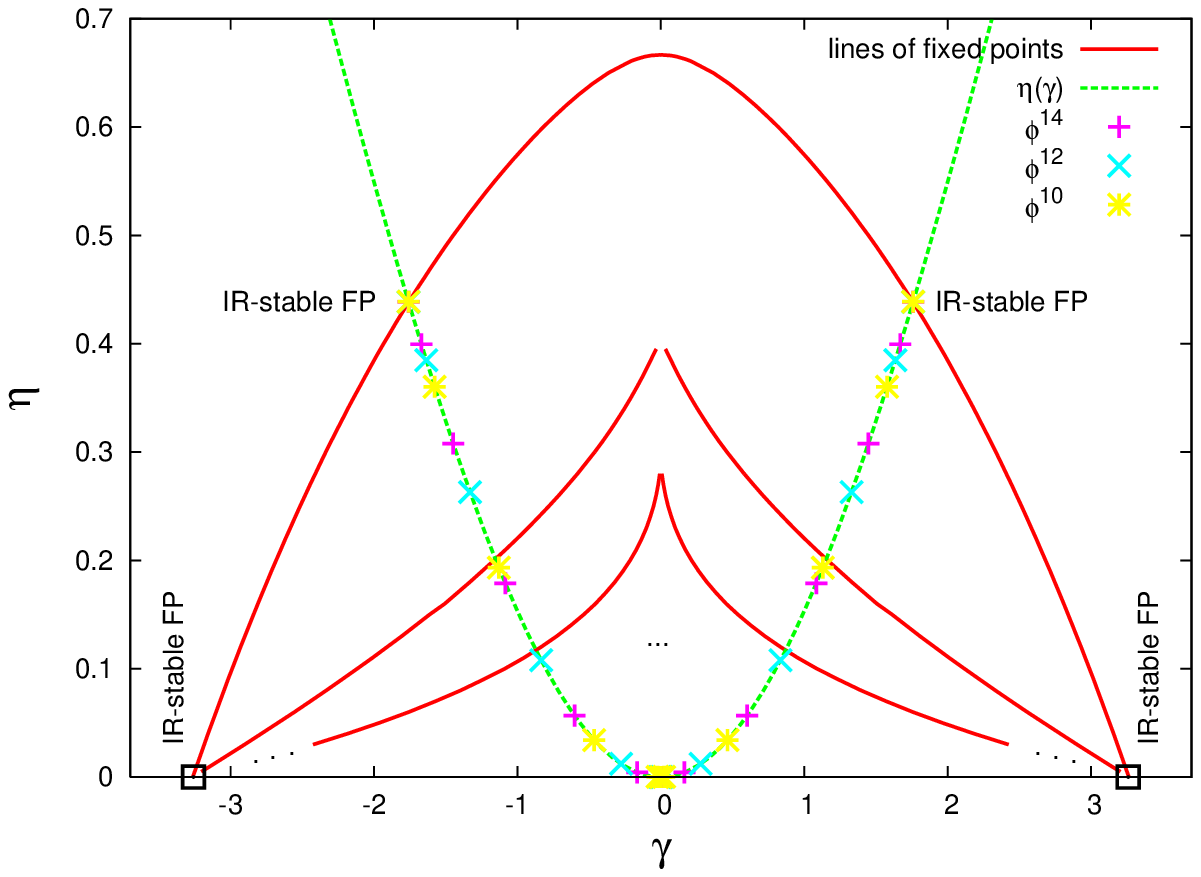}
\caption{Left panel: regular potentials for $\eta=0.1$. Their asymptotic
behavior is $\w''(\chi)\sim\chi^{18}$; Right panel: Lines of fixed points in the
	$\eta$-$\gamma$ plane (solid curves) and the anomalous dimension as a function of $\gamma=2\lambda$ obtained from Eq.
\eqref{eq:NextToLead4} (dotted curve). We also display the fixed point solutions
obtained from a polynomial approximation of Eq.~\eqref{eq:NextToLead1} and
\eqref{eq:NextToLead3} for different truncations.
\label{fig:RegPotential}}
\end{figure*} 

Now, for a non-vanishing $\eta$ we find a new class of solutions.  For these
new solutions, we consider again the derivative of
Eq.~\eqref{eq:NextToLead1}. This leads to the following fixed point equation
for $\ua=\w''_\ast$ (note that $\ua$ now contains a wave function
renormalization in contrast to section \ref{sec:FixedPointStructure}):
\begin{align}
&\frac{\ua''}{4\pi}\left[\left(\frac{\eta(3+\ua^2)}{2\ua^4}-\frac1{\ua^2}\right)
	\ln\left(1+\ua^2\right)+\frac{2}{1+\ua^2}-\frac{3\eta}{2\ua^2}
\right]\nonumber\\
	&=(\eta-1)\ua+\frac{\eta\chi}2\ua'+
	\frac{\ua'^2}{2\pi}\left[\frac{1+3\ua^2}{\ua(1+\ua^2)^2}-\frac{\eta(3+2\ua^2)}{\ua^3(1+\ua^2)}
	\right.\nonumber\\&\left.
\phantom{=}	+\left(
	\frac{\eta(6+\ua^2)}{2\ua^5}-\frac{1}{\ua^3}\right)\ln\left(1+\ua^2\right)
	\right].\label{eq:fp1}
\end{align}
For an initial condition in terms of an odd superpotential $W_\Lambda$ at the
UV scale $k=\Lambda$, $\w_t$ is odd and the fixed point solution
$\ua=\w_\ast''$ vanishes at the origin. Thus for weak fields we have
$\ua(\chi)\ll 1$ and we may expand the logarithm in powers of $\ua$. The
resulting fixed point equation for small $\ua$ is regular for $\ua\to 0$ and
reads
\begin{align}
&\nonumber
\frac{\ua''}{16\pi}\left((\eta-4)+(6-\eta)\ua^2+\frac{5}{6}(\eta-2)\ua^4 +\dots\right)\\
=&
-\frac{\ua\ua'^2}{8\pi}\left((6-\eta)+\frac{5}{3}(\eta-8)\ua^2
+\frac{21}{10}(10-\eta)\ua^4+\dots\right)\nonumber\\&-\frac{\eta}{2}\chi\ua'-(\eta-1)\ua.
\end{align}
Following \cite{Neves:1998tg}, we first consider $\eta$ in Eq. \eqref{eq:fp1} as
a free parameter.  The initial conditions $\ua(0)=0$ and
$\ua'(0)=\gamma=2\lambda$ are parameterized by the slope $\gamma$ at the
origin. A solution of Eq.~\eqref{eq:fp1} with generic slope will run into a
singularity because the factor multiplying $\ua''$ eventually becomes zero.  By
fine-tuning the slope it is possible to find regular solutions for a given
value of $\eta$.  In Fig.~\ref{fig:RegPotential} (left panel) we show three
regular potentials for $\eta=0.1$.  For large values of $\chi$, they behave like
$\w''(\chi)\sim\chi^{18}$.

These regular solutions define curves of fixed-point solutions in the
$\gamma$-$\eta$ plane. This is shown in Fig.~\ref{fig:RegPotential} (right
panel). For $\eta=2/3$ we find a potential that behaves as $\ua\sim\chi$ in the
asymptotic region. We do not find solutions with larger values for $\eta$.  For
$0<\eta<2/3$ it follows from simple monotony arguments that the factor
multiplying $\ufrak''$ in Eq.~\eqref{eq:fp1} has only one node at some value
$\phi_0$. The potentials are therefore regular if the right-hand side of Eq.
\eqref{eq:fp1} vanishes for the same value $\phi_0$, which is achieved by a
fine-tuning of $\gamma$.
The outermost curve in Fig.~\ref{fig:RegPotential} (right panel) corresponds to
a potential $\w''_\ast$ with no nodes, the next curve to potentials with one node and the
third curve to potentials with two nodes. We expect to find more curves for
small $\eta$ and $\gamma$ corresponding to potentials with more nodes. In
Fig.~\ref{fig:RegPotential} (right panel) we also display
$\eta(\gamma)=4\gamma^2/(\gamma^2+8\pi)$ obtained from equation
\eqref{eq:NextToLead3}.
 
\begin{table}
\begin{center}
\begin{ruledtabular}
\begin{tabular}{ccc}Number of nodes&$\eta$&${\nu_W}$\\\hline
0&0.4386&1.2809\\ 1&0.20	&1.11\\
2&0.12	&1.06\\
\end{tabular}
\end{ruledtabular}  
\caption{Critical exponents of the first fixed points.\label{tab:exponents}}
\end{center} 
\end{table}

The polynomial approximation to the fixed point solution of Eq.
\eqref{eq:NextToLead1} and Eq.~\eqref{eq:NextToLead3} converges to the
maximally IR-stable fixed point with $\eta=0.4386$ and
$\gamma=1.759$, which is just the point of intersection with the line
of fixed points corresponding to potentials with no nodes. The intersection
points of the $\eta(\gamma)$ line with the other lines of fixed points with
higher numbers of nodes then give estimates for the critical exponents of
these other fixed points at next-to-leading order in the derivative expansion.
E.g., we find $\eta\simeq0.20$ for the fixed point with one node and $\eta\simeq0.12$
for the fixed point with two nodes. The corresponding critical exponent
$\nu_W$ then follows directly from the superscaling relation
\eqref{eq:superscaling}. They are listed in Tab. \ref{tab:exponents}. The point
characterized by $\eta=0$ and $\gamma=3.529$ in Fig.~\ref{fig:RegPotential} 
(right panel) (see appendix \ref{sec:IRstableFP}) belongs to a solution of the
type discussed in section \ref{sec:FixedPointStructure}, where the maximally IR-stable solution can be extended without cusps beyond the critical value of the potential $\w''$.

\subsection{Synthesis: derivative expansion results at leading order and
  next-to-leading order} 
 
At a first glance, the fixed-point potentials obtained at the various orders
in the derivative expansion seem even qualitatively different. At
leading-order, we find oscillating solutions, solutions with cusps, and
solutions with a compact target space. By contrast, the next-to-leading-order
solutions admit superpotentials that can be extended to infinite field
amplitude with a standard powerlaw asymptotics $W'(\phi\to \infty)\sim
\phi^{2/\eta -1}\to \infty$. 

Of course, there are also many similarities, as the next-to-leading-order
fixed-point potentials can be classified by their number of nodes, i.e., they
typically exhibit an oscillating behavior for small fields. In addition, we
expect the occurrence of superpotentials with singular structures at finite
field values which have not been searched for as systematically as in the
leading-order case. 

The key to a unified understanding of both orders is provided by the anomalous
dimension $\eta$, as the new large-field asymptotics at next-to-leading order
is induced by a nonzero value for $\eta$. It should be kept in mind that we
deduce the nonzero value for $\eta$ from a small-field expansion of the full
flow of the wave function renormalization $Z_k(\phi)$. Beyond this expansion,
the anomalous dimension will acquire a field dependence $\eta\to
\eta(\phi)$. From the general form of the flow equation, we expect that the
large-field limit is characterized by $\eta(\phi\to\infty)\to 0$. We therefore
conjecture that the true large-field asymptotics of the fixed-point
superpotentials lies in-between the leading- and next-to-leading-order
results. More precisely, we expect that a standard asymptotic behavior
$W'(\phi\to \infty)\to \infty$ persists, but the powerlaw behavior
$\phi^{2/\eta -1}$ may be replaced by a stronger divergence. 

In any case, the rapid convergence of the polynomial expansion in both orders
of the derivative expansion, as well as quantitative agreement between
observables derived from the polynomial expansion and from the full solution
support the reliability of the overall picture arising from the derivative
expansion.

%======================================================
\section{The Gaussian Wess-Zumino model}
%======================================================
\label{sec:gaussian-fixed-point}

In principle, each of the fixed points defines a different UV completion of
the Wess-Zumino model and therefore a different physical system with a
different number of physical parameters. In the following, we concentrate on
the Wess-Zumino model defined at the Gau\ss ian fixed point corresponding to
an asymptotically free theory. At least seemingly, this is a natural choice,
as it has also often been used in lattice computations. However, the Gau\ss
ian fixed point actually has infinitely many relevant directions, as is
already revealed by perturbative power-counting. As a consequence, there are
strictly speaking infinitely many physical parameters. 

In practice, one usually starts with a classical superpotential including
quadratic perturbations of the Gau\ss ian fixed point,
$W'_\Lambda=\bar\lambda_\Lambda(\phi^2- \bar a_\Lambda^2)$, at the UV cutoff
$k=\Lambda$, implying that infinitely many couplings have been set to zero at
that scale. Since the RG trajectories are not regulator independent, it will
still be difficult to compare our results with those of, say, lattice
computations, as the same physical system with lattice regularization might
have a very different action at the lattice cutoff $\Lambda=\pi/a$. In fact,
we find a substantial quantitative regulator dependence for non-universal
quantities within the functional RG calculations; see App.~\ref{app:reg} which might imply that a
meaningful comparison with other methods should only be made on a
qualitative level. 

On the other hand, one may interpret the choice of the regulator as belonging
to the definition of the theory itself: the regulator together with the
initial condition in the form of a quadratic perturbation specifies the RG
trajectory uniquely also at a finite scale $\Lambda$. In the general case,
this viewpoint has the disadvantage that a change of the cutoff scale
$\Lambda$ on the line of constant physics generically involves an adjustment
of the couplings of infinitely many operators. In order to find out how much
these operators actually affect the flow of the superpotential, we have
varied the cutoff scale $\Lambda$. For a given $\Lambda$, we adjust the
couplings in $W_\Lambda'=\bar\lam_\Lambda(\phi^2-\bar a_\Lambda^2)$ such that we
obtain fixed reference couplings $a_{\Lambda_0}$ and $\lam_{\Lambda_0}$ at a
reference scale $\Lambda_0$, ignoring higher-order couplings. This way of
looking at the cutoff-dependence is very much motivated by similar procedures
in recent lattice simulations of the two-dimensional Wess-Zumino model in
\cite{Wozar:2008jb,Kastner:2008zc}. For large enough $\Lambda$, the solutions
of the flow equation show that the dependence of, e.g., the ground state
energy at $k=0$ on the actual cutoff scale is small. This observation helps
making the viewpoint of including the regulator in the definition of the
theory practicably applicable. Still, it has to be emphasized that we observe
a significant regulator dependence of the nonuniversal quantities, see
App.~\ref{app:reg}.

%======================================================
\subsection{Numerical solution of the  flow equation in local-potential
  approximation}
%======================================================
 \begin{figure*}
\includegraphics[width=.95\columnwidth]{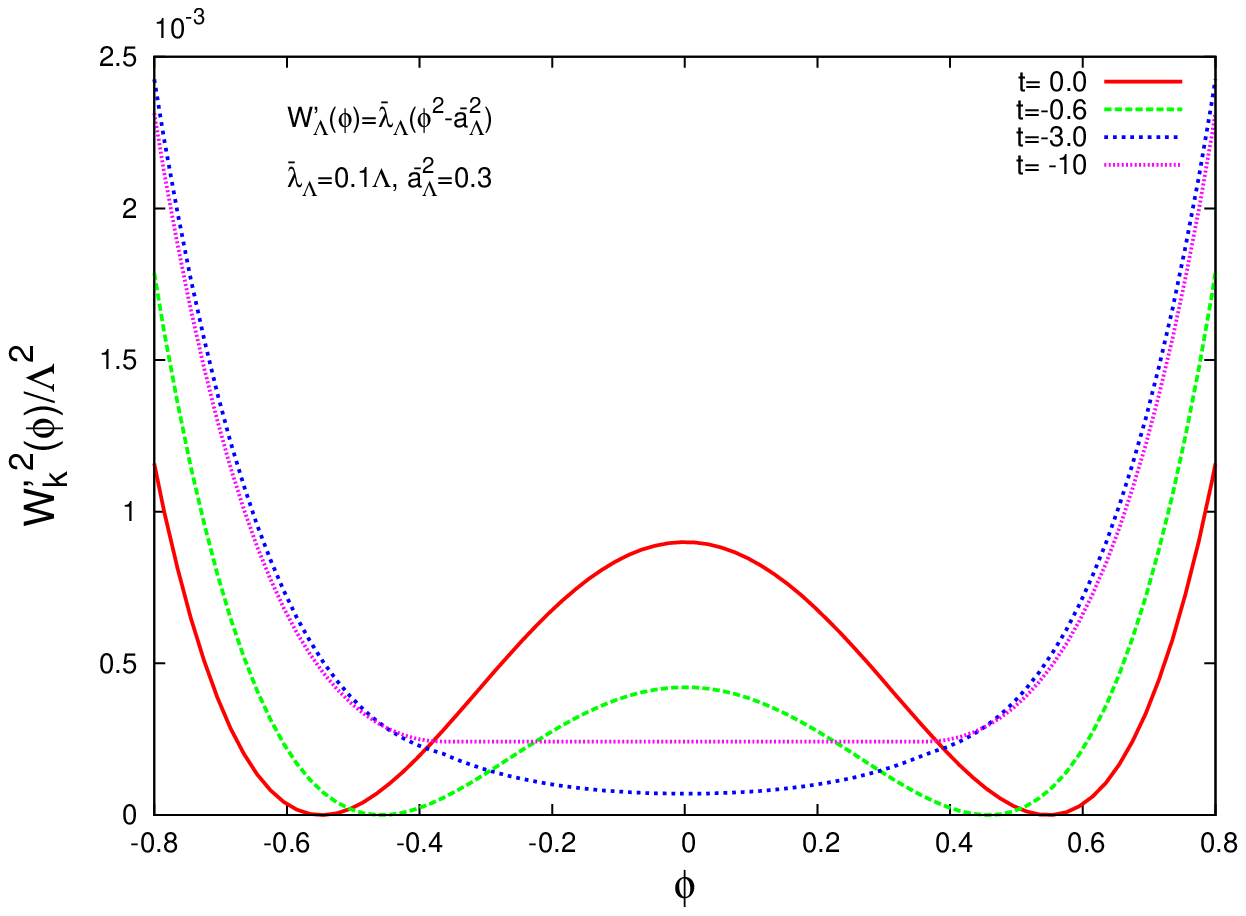}\hfill
\includegraphics[width=.95\columnwidth]{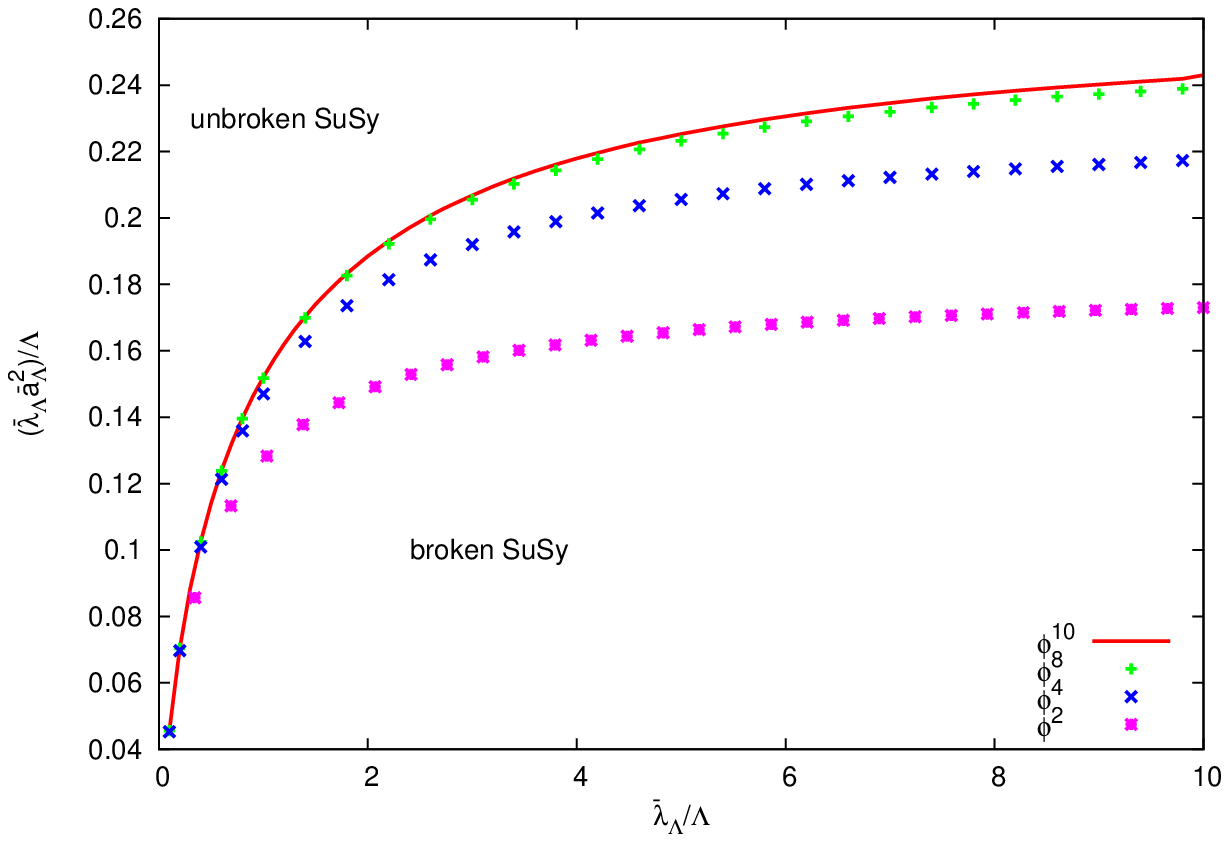}
\caption{Left panel: Flow of a potential with dynamical supersymmetry breaking, $W'_\Lambda(\phi)=\bar\lambda_\Lambda({\phi^2}-\bar a_\Lambda^2)$, $\Lambda =
 1000$, $\bar\lambda_\Lambda = 100$, $\bar a_\Lambda^2 = 0.3$; Right panel:
 Phase diagram in the space of the
dimensionless couplings specified at the cutoff scale  $\Lambda$ for different
truncations.\label{fig:FlussPotential}}
\end{figure*}

At the cutoff scale $\Lambda$, we start with
$W'_\Lambda=\bar\lambda_\Lambda(\phi^2- \bar a_\Lambda^2)$, where
$\bar\lambda_\Lambda\phi^2$ is a relevant perturbation of the Gau\ss ian fixed
point. For large values of the field, the right-hand side of
Eq.~\eqref{eq:flow3a} vanishes such that $\pa_kW_k'\simeq 0$ and the effective
potential remains close to the classical potential.  It follows that the
dimensionless potential $w_t=W_k/k$ diverges for large $\phi$ in the
infrared. On the other hand, we expect that the dimensionful
$W''(\phi)=ku_t(\phi)$ converges to zero for small fields, since $u_t$ can be
attracted by the maximally IR-stable fixed point and converges to a bounded
function $u$.

Because of the singularity in the fixed point equation for $u$, the solution
of the partial differential equation \eqref{eq:flow3a} poses a numerical
challenge. To meet this challenge, we exploit the fact that the polynomial
solutions with scale-dependent coefficients \eqref{eq:Pol1} yield excellent
approximations to the solution of the full partial differential equation for small
fields $\phi<\phi_{\rm crit}$. For the fixed-point solution, this has been
demonstrated earlier, see Fig.~\ref{fig:FitFixPoint}, left panel. Thus, we use
a polynomial approximation for $\vert \phi\vert<\phi_{\rm crit}$. 
For large fields $\abs{\phi}>\phi_{\rm crit}$, the polynomial approximation 
fails and we solve the partial differential equation numerically. 
As boundary conditions for the numerical solutions, we impose $W_k'(\phi)=W_\Lambda'(\phi)$ 
for $\vert\phi\vert\to\infty$ and $W_k(\phi_{\rm crit})$ equal to the polynomial 
approximation at $\phi_{\rm crit}$ and scale $k$.
Figure \ref{fig:FlussPotential} (left panel) shows the flow of such a potential.

%======================================================
\subsection{The phase diagram}
%======================================================

Depending on the parameters in $W'_\Lambda=\bar\lambda_\Lambda(\phi^2-\bar
a_\Lambda^2)$ we may end up with a broken or unbroken supersymmetry in the
infrared.  Note that $\lam_k$ is dimensionful, whereas $a_k$ is dimensionless.
The rescaled dimensionless couplings have an index $t=\ln k/\Lambda$, e.g.
$\lam_k=k\lam_t$ and $a_k=a_t$.  In the following, we determine the parameter
region for which supersymmetry is dynamically broken. A criterion for
supersymmetry breaking is provided by a nonvanishing ground state energy,
given by the minimal value of $V(\phi)=\frac12W'{\,^2}(\phi)$, where $ W'=
W'_{k\to 0}$. The ground state energy is nonzero if and only if $W'(\phi)>0$
for all $\phi$. Since $W'(\phi)$ is minimal at the origin we may use the
polynomial approximation for which $W_k'(0)=-\lam_k\cdot
a^2_k=-(k\lambda_t)\cdot a^2_t$.  The dimensionless coupling $\lambda_t$ flows
to the IR fixed-point value $\lambda^*$, whereas the coupling $a^2_t$ diverges
for $k\to 0$ or equivalently for $t\to-\infty$, such that the dimensionful
quantity $k\,\lambda_t\cdot a^2_t$ converges to a \emph{finite value}. This is
a direct consequence of the fact that the divergence of $a_t^2$ at the fixed
point is governed by the critical exponent $\theta^0=1$. The supersymmetric
phase is characterized by $a^2_t\to+\infty$ and a double well potential
$W'^{\,2}_k$, whereas the broken phase is characterized by $a^2_t\to-\infty$
and a single well potential. 
 
This gives rise to a strong analogy between dynamical supersymmetry breaking
in this and many other supersymmetric systems and quantum critical phenomena
in strongly correlated fermion systems \cite{Gies:2009az}: both phenomena are
governed by a control parameter of the (su\-per-)potential which plays the role
of a bosonic mass term; see, e.g., \cite{strack-2009} for an RG treatment of a
semi-metal--superfluid quantum phase transition in a fermionic system. This
control parameter, which is usually called $\delta$ in quantum critical
phenomena, is associated to the combination $\delta=
\bar\lambda_{\Lambda}\cdot \bar a^2_{\Lambda}/\Lambda$ in our case. The
critical value $\delta_{\text{cr}}$ of the control parameter marks a quantum
critical point of a quantum phase transition.

In Fig.~\ref{fig:FlussPotential} (right panel), we depict the phase diagram in
the space of dimensionful couplings $\bar\lambda_{\Lambda}$ and
$\bar\lambda_{\Lambda}\cdot \bar a^2_{\Lambda}$ in units of the UV cutoff
$\Lambda$ for increasing truncation order. 
In the $\phi^2$ truncation, the system of differential equations reads
(see equation \eqref{eq:Pol1})
\begin{align}
		\partial_t a_t^2
		&=\frac{1}{2\pi }-\frac{6\lambda_t^2\cdot a_t^2}{\pi },\quad
	\partial_t\lambda_t=-\lambda_t+\frac{6\lambda_t^3}{\pi},
	\label{eq:phase1}
\end{align}
which can be solved analytically. The phase transition curve in the
$(\bar\lam_\Lambda \bar a^2_\Lambda,\bar\lam)$ plane is given by the
initial values for which $ a^2_{t\to-\infty}$ changes sign. 
This condition yields
\begin{align}
\frac{\bar\lambda_\Lambda
\bar a^2_{\Lambda}}{\Lambda}=\frac{\arcsin (\al)}
{\sqrt{24\pi}\,\al},\qquad
\al^2=1-\frac{\pi \Lambda}{6\bar\lam_{\Lambda}}.
\label{eq:phase3}
\end{align}
In order to find the approximate phase transition curve for the higher order
polynomial truncations, we have integrated the corresponding systems of flow
equations numerically. From the $\phi^2$ to the $\phi^4$ truncation, the
transition curve moves considerably upwards because the coupling $b_{4,t}$
enters the flow equation for $a^2_t$ in Eq.~\eqref{eq:Pol1}. The higher
couplings $b_{6,t},b_{8,t},\dots$ enter the differential equation for $a_t$
only indirectly, and, as a result, the approximate phase transition curves
converge rapidly with increasing order of the truncation. We find that there
exists a critical value for the cutoff parameter $\bar\lambda_\Lambda \bar
a_\Lambda^2|_{\text{crit}}$ characterizing the phase transition in the strong
coupling limit $\bar\lambda_{\Lambda}\to\infty$. To lowest order,
Eq.~\eqref{eq:phase3} leads to a critical value of $\bar\lambda_{\Lambda} \bar
a_{\Lambda}^2 |_{\text{crit}}/\Lambda=\sqrt{\pi/96}\simeq 0.181$. Our best
estimate for this critical value derived from a numerical higher-order
solution is $\bar\lambda_{\Lambda} \bar a_{\Lambda}^2
|_{\text{crit}}/\Lambda\simeq0.263$. We conclude that supersymmetry can never be
broken dynamically above this critical value. This agrees qualitatively with
earlier results in the literature \cite{Witten:1982df, Beccaria:2004pa}.

A more quantitative comparison to lattice simulation is inflicted by the
strong regulator dependence of nonuniversal quantities, such as the bare
critical coupling values discussed above, see also App.~\ref{app:reg}.  For
instance in \cite{Beccaria:2004ds,Beccaria:2004pa}, the phase diagram and the
ground state energy of the present model were investigated anew.  For a fixed
$\bar\lambda_\Lambda/\Lambda=0.5$, a phase transition from a state with broken
to a state with unbroken supersymmetry was observed at the critical value $
(\bar\lambda \bar a^2)|_{\text{crit}}/\Lambda=0.48$. With two different
methods they obtained $0.40$ and $0.52$ for this critical value but they state
$0.48$ to be the most solid value. In these works, the thermodynamic limit has
been performed without an accompanying continuum extrapolation. On the other
hand, the quoted numbers for $(\bar\lambda \bar a^2)_\Lambda$ were in reasonable good
agreement with earlier results obtained with the help of the worldline path
integral method \cite{Beccaria:2003ba}. As demonstrated in App.~\ref{app:reg},
a direct comparison of bare quantities between the present work and
\cite{Beccaria:2004ds,Beccaria:2004pa} is anyway not meaningful due to the
scheme dependence. If more lattice points in the phase diagram were available,
dimensionless coupling ratios could be compared which are likely to be less
affected by scheme dependencies.

%======================================================
\subsection{Mass scaling}
%======================================================

The scale-dependent bosonic mass can be read off from the bosonic potential in
the on-shell formulation, which is given by $\bar{V}_k(\phi)=\frac{1}{2} (W_k'(\phi))^2$
in our truncation. In terms of renormalized fields $\chi=Z_k \phi$, the
dimensionful renormalized bosonic potential and bosonic mass thus is
\begin{equation}
V_k(\chi)= \frac{1}{2} \, \frac{1}{Z_k^2}\, \big(W_k'(\chi/Z_k)\big)^2, \quad
m_k^2 = V_k''(\chi_{\text{min}}), 
\end{equation}
where $\chi_{\text{min}}$ denotes the minimum of the effective potential
$V_k(\chi)$. The true mass can then be read off in the limit $m=\lim_{k\to 0}
m_k$. 

In the broken phase, the minimum of both $V_k$ and $W_k'$ is at $\chi=0$, such
that the bosonic mass yields
\begin{equation}
m_k^2 = \frac{1}{Z_k^4} \, W_k'(0) W_k'''(0) = 2 k^2\, \lambda_t^2 |a_t^2|,
\end{equation}
where we have used the fact that the renormalized parameter $a_t^2$ is
negative. Assuming that the system is dominated by the maximally IR-stable
fixed point with $\lambda_t \to \lambda^\ast$ and $a_t^2\sim k^{-1/\nu_W}$,
the renormalized mass scales as
\begin{equation}
m_k^2 \sim k^{1+ \frac{\eta}{2}} 
\end{equation}
where we have employed the superscaling relation \eqref{eq:superscaling}. For
$\eta>-2$, the renormalized bosonic mass scales to zero upon attraction of the
maximally IR-stable fixed point. Indeed, for the Gau\ss ian Wess-Zumino model
considered here, we observe that the flow in the broken phase is always
attracted by the maximally IR-stable fixed point. Together with the fact that
the broken phase also goes along with a massless goldstino, we conclude that
the broken phase remains massless in both degrees of freedom. 

In a certain sense, the underlying limit $k\to0$ represents an extreme point
of view. Any experiment as well as any lattice simulation will involve an IR
cutoff scale $k_{\text{m}}$ characterizing the measurement, e.g., the scale of
a momentum transfer, the detector size or the lattice volume. Any measurement
therefore is not sensitive to $k\to 0$ but to $k\to k_{\text{m}}>0$. We
conclude that any measurement of the bosonic mass in the broken phase will
give a nonzero answer proportional to the measurement scale, whereas the
goldstino will be truly massless.  

In the broken phase, the superscaling relation also has a special consequence
for the unrenormalized potential. We observe that
\begin{equation}
W'(0) = -\bar{\lambda}_k \bar{a}_k^2 = -kZ_k \lambda_t a_t^2 \sim
k^{1-\frac{\eta}{2}} \, k^{-1/\nu_W} \to \text{const.}, 
\end{equation}
where we have used Eq.~\eqref{eq:superscaling} and $Z_k\sim k^{-\eta/2}$. Therefore, the
flow of the superpotential freezes out near the origin if the system is in the domain of
attraction of the maximally IR-stable fixed point.

\begin{figure}
\centering{ 
\includegraphics[width=.9\columnwidth]{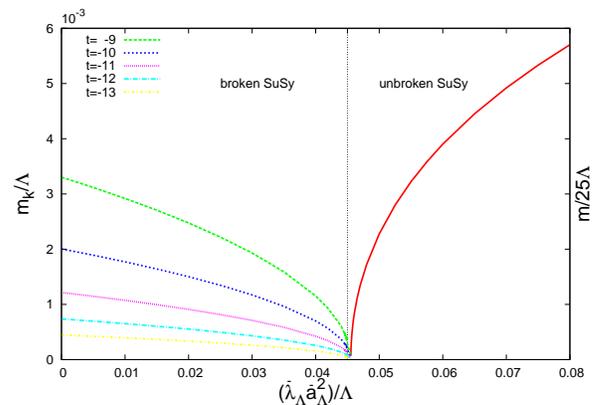}}
\caption{Renormalized mass at different scales $k$ as a function of the
  initial condition $\bar\lambda_\Lambda \bar a^2_\Lambda$ at an initial
  coupling of $\bar\lambda_\Lambda=0.1$. \label{fig:Masse}} 
\end{figure}  

In the supersymmetric phase, the bosonic as well as the fermionic mass is
given by 
\begin{equation}
m_k^2 = \frac{1}{Z_k^4} \, \big(W_k''(\chi_{\text{min}}/Z_k)\big)^2,
\end{equation}
where we made use of the fact that $W_k'$ vanishes at the minimum. For typical
flows, $W_k''$ remains positive at the minimum also in the $k\to 0$
limit. This leads to a generic decoupling of the massive modes, once $k$ drops
below the mass threshold. For an accurate inclusion of this mass decoupling,
also the anomalous dimension needs to be evaluated at the minimum
$\chi_{\text{min}}$, implying $\eta\to 0$ for $k^2\ll m_k^2$,
cf. Eq.~\eqref{eq:NextToLead3}.

For simplicity, we confine ourselves to the local-potential approximation and
set $Z_k\to1$, $\eta\to 0$ for a computation of the masses across the phase
transition. Our results are displayed in Fig.~\ref{fig:Masse} as a function of
the initial-condition parameter $\bar\lambda_\Lambda \bar a^2_\Lambda$ of the
relevant direction. The calculation has been performed at an initial value for
the coupling of $\bar\lambda_{\Lambda}/\Lambda=0.1$. We observe a critical
value of $\bar\lambda_\Lambda \bar
a^2_\Lambda|_{\text{crit}}/\Lambda\simeq 0.045,$ above which supersymmetry
remains unbroken and the theory is massive. Both masses vanish at the quantum phase transition. Below the
critical value, supersymmetry is broken and the goldstino is massless. The
bosonic mass also approaches zero for $k\to 0$, but remains finite for any
finite value of $k$, potentially representing a measurement scale.

 %======================================================
\section{Conclusions}
%======================================================
In this work, we have constructed a manifestly supersymmetric flow equation for
the application of the functional RG to the two dimensional $\mathcal N=1$
Wess-Zumino model. The regularization turns out to be similar to the one in
supersymmetric quantum mechanics.  This is not surprising since supersymmetric
quantum mechanics can be derived from the Wess-Zumino model by a dimensional
reduction.

For approximate solutions to the flow equation, we have employed an
expansion of the effective action in terms of field operators containing
increasing powers of supercovariant derivatives. This provides a systematic
approximation scheme that preserves supersymmetry.  We obtain a flow equation
for the superpotential and, at next-to-leading order, for the wave
function renormalization.

The Wess-Zumino model has a highly nontrivial fixed-point structure in terms
of scaling solutions for the superpotential. At leading- as well as
next-to-leading order in the derivative expansion, we find fixed-point
superpotentials which can be classified by an increasing number of relevant
directions. At leading order, the classification of fixed-point solutions can
largely be done on analytical grounds. Here, we find oscillating
sine-Gordon-type or even bouncing solutions on the one hand, and solutions
confining the field to a compact target space on the other hand. 
For the regular oscillating solutions, the large-field asymptotics is turned
into a standard form $W'(\phi\to\infty)\to\infty$ at next-to-leading order,
where the fixed-point superpotentials can be classified also according to
their number of nodes. This is reminiscent to fixed points of the effective
potential in two-dimensional bosonic theories which can be related to
conformal field theories  \cite{Morris:1994jc,Neves:1998tg}. Exploring the
connection between the present supersymmetric models at their fixed points and
conformal field theories remains an interesting question for future work.
 
Each fixed point defines its own universality class, the physics of which is
determined by the RG relevant directions of the fixed points. In other words,
each fixed point defines a different Wess-Zumino model with the number of
physical parameters given by the number of relevant directions. One extreme is
provided by the Gau\ss ian fixed point which has infinitely many relevant
directions in agreement with perturbative power-counting arguments. The other
extreme is given by the maximally IR-stable fixed point which has only one
relevant direction. We have demonstrated that this relevant direction is
exactly given by $a_t^2$ and is shared by all other fixed points as well. 

At leading- and next-to-leading order, the critical exponent associated with
this relevant $a_t^2$ direction can be computed exactly. At next-to-leading
order, this yields an intriguing relation between this critical exponent of
the superpotential $\nu_W=1/\theta^0$ and the anomalous dimension $\eta$ of
the field. As such a relation between critical exponents associated with
correlation functions is not known from Ising-type systems (where
thermodynamical exponents are related by scaling relations among each other
and by hyperscaling relations to the correlation exponents), this {\em superscaling}
relation appears to be a unique property of supersymmetric theories. 

The superscaling relation has a direct consequence for the renormalized superpotential
$W$ at the origin in field space. It dictates a freeze out of the derivative
of the superpotential at the origin in the deep IR, if the system is governed
by one of the fixed points. 

As an example for a model with dynamical supersymmetry breaking, we considered
the Wess-Zumino model defined by a quadratic perturbation of the Gau\ss ian
fixed point. In addition to the initial coupling value $\lambda|_\Lambda$, the
model has a control parameter $\lambda a^2|_\Lambda$ the value of which
decides about the realization of supersymmetry by the ground state of the
theory. At a given coupling, supersymmetry can only be broken for $\lambda
a^2|_\Lambda$ below a certain critical value which is in accord with a general
argument by Witten. We have computed the critical line of quantum phase
transitions in the coupling--control-parameter plane for a wide range from
weak to strong coupling. Most importantly, the control parameter stays finite
even at arbitrarily large coupling. 

We have also computed the masses of the lowest fermionic and bosonic
excitations across the quantum phase transition. In the supersymmetric phase,
both masses are equal and nonzero, but drop to zero at the phase
transition. In the broken phase, the fermion has a massless goldstino mode. We
observe that the boson also becomes massless in the broken phase in the deep
IR, $k\to 0$, but stays finite for any finite $k$. Associating $k$ with a
typical measurement momentum scale (say, inverse length of a detector), our
results predict that the bosonic mass in the broken phase is proportional to
the momentum scale set by the detector.

The critical properties of the quantum phase transition remain an interesting
open problem. In this work, we have considered the ground state energy or the
bosonic and fermionic masses of the theory as order parameters for the
symmetry. In our numerical results, we have found no hint for typical scaling
behavior of these quantities near the phase transition so far. Actually, a
true field-valued order parameter is given in terms of the expectation value
of the auxiliary field $F$. It is therefore natural to expect that
order-parameter fluctuations of the $F$ field play an important role near
criticality and eventually establish a scaling behavior. Technically, this
requires the inclusion of potential terms for the $F$ field. As these appear at
higher orders in the supercovariant derivative expansion, a quantitative
description of the critical regime remains a technical challenge.

From the perspective of the functional-RG tool box, we have solved the flow
equation for the superpotential both in a polynomial expansion as well as with
a full numerical solution of the corresponding partial differential
equation. Whereas the polynomial approximation for the potential is only
reliable in the vicinity of its expansion point with a finite radius of
convergence, it often suffices to extract reliable quantitative information
about physical observables such as critical exponents or the phase diagram.

In the context of the phase diagram of the Gau\ss ian Wess-Zumino model, we
have expressed our concern that a quantitative comparison with other methods
such as lattice simulations can be plagued by the fact that the model has
infinitely many relevant directions. As a direct consequence, the results of
the model defined by a certain (say quadratic) perturbation of the fixed point
at a fixed UV scale $\Lambda$ are regulator dependent, as this defining
initial condition is not universal. A much better comparison could arise from
defining a Wess-Zumino model, e.g., in the vicinity of the next-to-maximally IR-stable
fixed point, where there are only two relevant directions and thus two tunable
physical parameters. Fixing these parameters in terms of two observables,
all other quantities are a universal scheme-independent prediction of the
theory.

%======================================================
\acknowledgments{Helpful discussions with G.~Bergner, C.~Wozar, T.~Fischbacher
and T.~Kaestner are gratefully acknowledged. FS acknowledges
support by the  Studienstiftung des deutschen Volkes. This
work has been supported by the DFG-Research Training Group 
''Quantum-and Gravitational Fields'' GRK 1523/1, the DFG grants Wi 777/10-1 and 
Gi 328/5-1 (Heisenberg program).}

\appendix
%======================================================
\section{Flow equation with wave function renormalization}
%======================================================
\label{sec:flowWithWaveFkt}
At next-to-leading order in the derivative expansion, a field-independent wave
function renormalization is included in the truncation via 
\begin{multline}
 \Gamma_k[\phi,F,{\psib},\psi]
 = \int d^2x
  \left[Z_k^2\left(\ft12\partial_\mu\phi\partial^\mu\phi+
	\ft{i}{2}\psib\slashed{\partial}\psi
	-\ft12F^2\right)\right.\\+\left.\ft12W_k''(\phi)\psib\gamma_\ast\psi-W_k'(\phi)F\right].\label{eq:wfr1}
\end{multline}
The cutoff action reads
\begin{equation}
\Delta S_k=\ha\int (\phi,F) Z^2_kR^{\rm B}_k\, {\phi \choose F}
+\ha\int dx^2\psib Z^2_k R^{\rm F}_k\psi,\label{eq:wfr3}
\end{equation} 
with $R^{\rm B}_k$ and $R^{\rm F}_k$ given  in Eq.~\eqref{eq:lpa13}.
The flow equation for the superpotential is obtained by a projection onto the
terms linear in the auxiliary field and by integration with respect to $\phi$. 
We obtain a similar flow equation as in Eq.~(\ref{eq:flow1}),
\begin{equation}
\begin{split}
\partial_kW_k(\phi)&=\ha\int \frac{d^2
p}{4\pi^2}\,\frac{(1+r_2)Z^2_k\,\pa_k(r_1Z^2_k) }{\Delta}\\&-\ha\int \frac{d^2
p}{4\pi^2}\,\frac{ (W''_k(\phi)+r_1 Z^2_k)\pa_k(r_2 Z^2_k)}{\Delta}.
\end{split}
\end{equation}
Including the wave function renormalization, the expression in the denominator
reads
\begin{equation}
\Delta=Z^4_k\,p^2(1+r_2)^2+(W_k''+r_1Z^2_k)^2.
\end{equation}
For the flow of the wave function renormalization, we project the flow
equation onto the terms quadratic in the auxiliary field.  As we consider only
a field-independent wave function renormalization, we can also project onto
$\phi=0$
\begin{align}\nonumber
	&\partial_kZ^2=-W_k'''(\phi)^2 Z_k^2
\int \frac{d^2 p}{4\pi^2}\,(1+r_2)\times\\&\Bigg[
	\frac{2Z_k^2 \left(W_k''(\phi )+r_1
   Z_k^2\right)(1+r_2)}{\Delta^3}\,\partial_k(r_1Z_k^2)  \\&
 +\,
   \frac{Z_k^4 p^2
   (1+r_2)^2-\left(W_k''(\phi )+r_1
   Z_k^2\right)^2}{\Delta^3}\,\partial_k(r_2Z_k^2)\Bigg]_{\phi=0}.\nonumber
\end{align}
As regulator shape functions, we choose $r_1=0$ and
$r_2=(k^2/p^2-1)\theta(1-p^2/k^2)$ for which the momentum integrals in the
flow equations can be calculated analytically. We obtain
 \begin{align}
\partial_k W_k=&
-\frac{{W''_k}^2\partial_k(k^2Z_k^2)+k^4Z_k^4\pa_k Z^2_k}{8\pi {W''_k}^3}
\ln\left(1+\frac{{W_k''}^2}{k^2Z_k^4}\right)\label{eq:wfr9}\nonumber\\
&+\frac{k^2 \partial_k Z^2_k}{8\pi W_k''}\\
\partial_k Z^2_k=&\frac{k}{4\pi}\left(\frac{Z_k W'''_k}{{W''_k}^2}\right)^2
\left[\frac{{W''_k}^2k\partial_k Z_k^2}{{W_k''}^2+k^2Z_k^4}\right.\label{eq:wfr11}\\
&\left.-k\partial_k Z_k^2\ln \left(1+\frac{{W_k''}^2}{k^2Z_k^4}\right)
-\frac{2
  {W''_k}^4Z_k^2}{({W_k''}^2+k^2Z_k^4)^2}\right]_{\phi=0}\!\!\!\!\!\!\!\!\!\!.
  \quad
\nonumber
\end{align}	
For the renormalized fields, $\chi=Z_k\phi$, the superpotential scales as
$\W_k(\chi)=W_k(\phi)$ and $\W_k'=W_k'/Z_k$, $\W_k''=W_k''/Z_k^2$, \ldots
 In
terms of the anomalous dimension $\eta=-\partial_tZ^2_k/Z_k^2$, the preceding
flow equations read
\begin{align}
k\partial_k \W_k(\chi)=&
\frac{\eta}{2}\chi\W'_k
-\frac{\eta\, k^2}{8\pi\W''_k}\\&+
\frac{(\eta-2)k^2{\W''_k}^2+\eta k^4}{8\pi
{\W''_k}^3}\ln\left(1+\frac{{\W_k''}^2}{k^2}\right),\nonumber\\
\eta=&\frac{k^2}{4\pi}\left(\frac{\W'''_k}{{\W''_k}^2}\right)^2
\left[\frac{\eta {\W''_k}^2}{{\W_k''}^2+k^2}\right.\\&\left.
-\eta\ln \left(1+\frac{{\W_k''}^2}{k^2}\right)
+\frac{2
{\W''_k}^4}{({\W_k''}^2+k^2)^2}\right]_{\phi=0}\!\!\!\!\!\!\!\!\!\!.\nonumber
\end{align}
In terms of the dimensionless superpotential $\w(\chi)=\W(\chi)/k$, this reads 
\begin{align}
\partial_t \w_k(\chi)=&
\frac{\eta}{2}\chi\w'_k-\w_k
-\frac{\eta}{8\pi\w''_k}\nonumber\\&+
\frac{(\eta-2){\w''_k}^2+\eta}{8\pi
{\w''_k}^3}\ln\left(1+{\w_k''}^2\right),\label{eq:wfr17}\\
\eta=&\frac{1}{4\pi}\left(\frac{\w'''_k}{{\w''_k}^2}\right)^2
\left[\frac{\eta {\w''_k}^2}{{\w_k''}^2+1}\right.\nonumber\\&\left.
-\eta\ln \left(1+{\w_k''}^2\right)
+\frac{2 {\w''_k}^4}{({\w_k''}^2+1)^2}\right]_{\phi=0},\label{eq:wfr19}
\end{align}
which agrees with Eqs.~\eqref{eq:NextToLead1} and \eqref{eq:NextToLead3}.

 \label{sec:strongReg1}

%======================================================
\subsection{Polynomial approximation }
%======================================================

\label{sec:strongReg3}
Next, we perform a polynomial expansion of the superpotential flow
Eq.~\eqref{eq:wfr17} including wave function renormalization.  We use the
conventions
$\w'_t(\chi)=\lambda_t(\chi^2-a^2_t)+\sum_{n=1}^Nb_{2n,t}\chi^{2n}$, leading
to the following system of coupled equations:
\begin{align}
\eta_t  =&\frac{(4 - \eta_t) \lambda_t^2}{2 \pi}\label{eq:polyeta}
\\
\partial_t\abar_t^2=&\frac{1}{8\pi }(4-\eta_t)-\eta_t\abar^2_t\nonumber\\&
		-\frac{\abar_t^2}{\lambar_t}
		\left(\frac{3 \left(\lambar_t^3-\bbar_{4,t} \right) }{\pi }
		-\frac{\left(2 \lambar_t^3-3 \bbar_{4,t}
   		\right) \eta_t}{4 \pi}\right)\nonumber\\
	\partial_t\lambar_t=&-\lambar_t+\frac{3 \left(\lambar_t^3-\bbar_{4,t}\right) }{\pi }
		-\frac{\left(2 \lambar_t^3-3 \bbar_{4,t}\right) \eta_t}{4 \pi }
		+\frac{3}{2}\eta_t\lambar_t\nonumber
   		\\
   \partial_t \bbar_{4,t}=&-\bbar_{4,t}-\frac{5  ( 8-\eta_t) \lambar_t^5-15 
   		\bbar_{4,t} (6-\eta_t) \lambar_t^2
	  }{3 \pi }\label{eq:polyCoupling}\\&-\frac{ 45 \bbar_{6,t} (4-\eta_t)
	  }{24 \pi }+\frac{5}{2}\eta_t\bbar_{4,t}\nonumber
   		\\
   \partial_t\bbar_{6,t}=&-\bbar_{6,t}-	\frac{7\bbar_{8,t} (4-\eta_t)
   		 }{2 \pi }-	\frac{
   		70 \bbar_{4,t} (8-\eta_t) \lambar_t^4 }{3 \pi }\nonumber\\
   		&+\frac{ 28 (10-\eta_t)
     	\lambar_t^7 }{5 \pi }%\nonumber\\&\phantom{-\bbar_{6,t}\;\;}
     	+\frac{ 
   		7( 3 \bbar_{6,t}  \lambar_t^2 +4 \bbar_{4,t}^2\lambar_t)(6-\eta_t)}{2 \pi }
    	\nonumber\\&+\frac{7}{2}\eta_t\bbar_{6,t} \nonumber\\
\vdots\nonumber 
\end{align}
Again, the first-order coupling $a_t$ does not influence the higher-order
flows.  Setting the left-hand side to zero, we find a nonlinear system of
algebraic equations for the fixed-point couplings. The solutions determine the
coefficients of the fixed-point superpotential in the polynomial expansion.

For a truncation at $2n$th order, we again find $2n+1$ real fixed points.  As
proved in the main text, each fixed point has at least one relevant direction
which is provided by the $a_t^2$ direction. The maximally IR-stable fixed
point has only this relevant (IR-unstable) direction. The other fixed points
can be classified according to their increasing number of further relevant
directions. The Gau\ss ian fixed point remains fully UV attractive also at
this order of the derivative expansion.

%======================================================
\section{Regulator dependence}
%======================================================
\label{app:reg}

\begin{table}
\begin{center}
\begin{footnotesize} 
\begin{ruledtabular}
\begin{tabular}{cccccccc}
&  \multicolumn{7}{c}{coefficients at IR-fixed point}\\\cmidrule(rl){2-8}
$2n$&$\lambar^\ast$&$\bbar_4^\ast$&$\bbar_6^\ast$&$\bbar_8^\ast$&$\bbar_{10}^\ast$&$\bbar_{12}^\ast$&$\bbar_{14}^\ast$
\\\hline 
2&1.023\\
4& 1.405& 1.301\\
6& 1.540& 2.040& 3.097\\
8& 1.593& 2.374& 4.868& 8.651\\
10&1.615& 2.523& 5.725& 13.38&26.00\\
12&1.625& 2.590& 6.124& 15.70&39.55& 81.19\\
14&1.629& 2.620& 6.308& 16.79&46.11& 121.8& 259.4\\
\end{tabular}
\end{ruledtabular}
\end{footnotesize}
\caption{Coefficients of the first few couplings in a polynomial expansion of
  the fixed-point superpotential at the maximally IR stable fixed-point for
  different truncations. The expected nonuniversal deviations from
  Tab.~\ref{tab:ExponIRstabilA} are due to the use of a different regulator.
\label{tab:CoeffIRStable}}
\end{center} 
\end{table} 

\begin{figure*}
\centering{\includegraphics[width=.9\columnwidth]{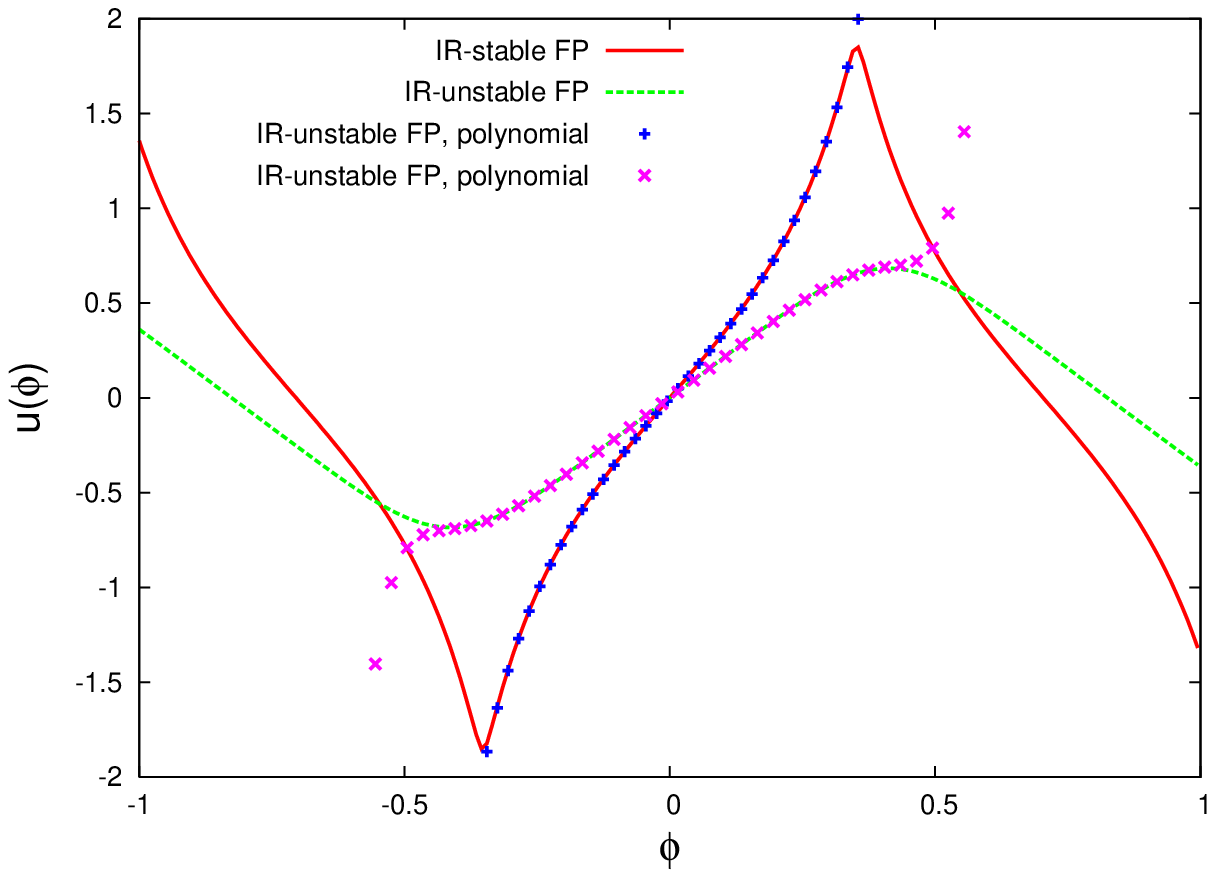}\hfill
\includegraphics[width=.9\columnwidth]{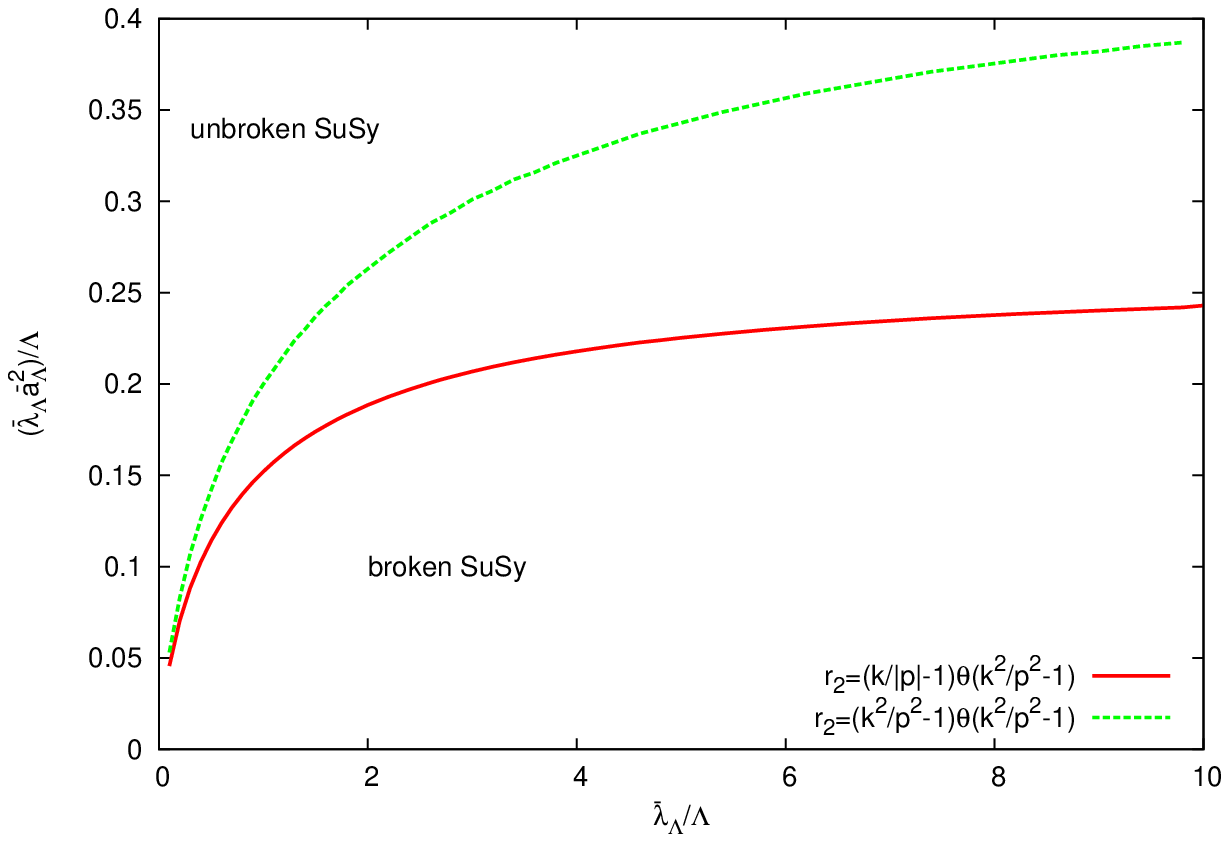} }
\caption{\label{fig:FPStrongReg} \emph{Left panel:} fixed-point
  superpotentials calculated from Eq.~\eqref{eq:flowStrong7} with
  $\lambda=1.6295$ for the maximally IR-stable fixed-point and
  $\lambda=1.0411$ for an IR-unstable fixed-point. \emph{Right panel:}
  Comparison between the critical bare values for $(\lambda a^2)_{t=0}$
  characterizing the phase transition calculated with the regulators
  $r_2=(k/p-1)\theta(p^2/k^2-1)$ and $r_2=(k^2/p^2-1)\theta(p^2/k^2-1)$. The
  difference is attributed to a strong scheme dependence of this nonuniversal
  quantity. }
\end{figure*}

\begin{table*}
\begin{center}
\begin{ruledtabular}
\begin{scriptsize}
\begin{tabular}{c|cccccccc}
 $\lambda^\ast$&\multicolumn{8}{c}{ Critical exponents $\theta^I$
 }\\\hline
$\pm 1.6315$&$-1.31$&$ -7.10$&$ -19.3$& $-42.7$&$ -84.8$&$ -158$&$ -285$&$-522$\\
 $\pm 1.4399$&5.43&$-1.49$& $-10.1$&$ -28.2$&$ -61.8$&$ -122$& $-227$&$-426$\\
 $\pm1.1463$& 19.5& 4.07&$ -1.51$&$ -11.8$&$ -33.9$& $-75.7$&$ -152$&$-298$\\
 $\pm.81753$& 28.1& 12.4& 3.23& $-1.39$& $-12.5$&$ -37.3$& $-85.7$&$-182$\\
 $\pm.49584$& $20.4 + 3.09 i$& 
$ 20.4 - 3.09 i$& 8.06& 2.54& $-1.16$&$ -12.5$&$-38.9$&$-95.0$\\
$ \pm.22322$& $11.9 + 8.85 i$& 
 $11.9 - 8.85 i$& 8.69& 5.07& 1.96& $-0.859$&$ -12.0$&$-39.8$\\
 $\pm.04903$& $4.27 + 1.14 i$& 
 $4.27 - 1.14i$ & $2.91 + 6.63 i$& $2.91 - 6.63 i$& 2.84& 1.47& $-0.547$&$-11.1$\\
 $\pm0.00042$& $1.57 + 0.125 i$& 
$ 1.57 - 0.125 i$&$1.43 + 0.70 i$&$ 1.43 - 0.703i$& 1.14& $0.542 + 0.982 i$&$ 0.542 - 0.982 i$
 & $-0.222$
 \\
 0&1& 1&1& 1& 1& 1& 1& 1
\end{tabular}
\end{scriptsize}
\end{ruledtabular}
\caption{Critical exponents $\theta^I$ (negative eigenvalues of the stability
  matrix) for a polynomial truncation at $2n=16$ for the nine different fix
  points in the local-potential approximation.
 The first exponent $\theta^0=1$ which is common to all fixed points is not
 shown here. This table should be compared with
 Tab.~\ref{tab:EigsStabMatrixN16} which has been computed with a different
 regulator. For all positive critical exponents, we find a remarkable degree
 of universality, as these exponents for the different regulators differ if at
 all by at most 10\%. The subleading negative critical exponents show larger
 variations. 
\label{tab:EigsStabMatrixN16StrongReg}}
\end{center}
\end{table*}

In this appendix, we repeat several studies of the local-potential
approximation as done in the main text but now for a different regulator shape
function $r_2=(k^2/p^2-1)\theta(p^2/k^2-1)$. This is actually equivalent to
the regulator used for at next-to-leading order in the main text. On the one
hand, this regulator comparison gives a rough estimate of how nonuniversal
quantities may vary. On the other hand, it supports all structural results of
the main text, which should in any case be universal.

 The flow equation for the superpotential \eqref{eq:flow1} now takes the form
\begin{align}
\partial_kW_k(\phi)=\frac{k }{4
   \pi  W''_k(\phi )}\ln \left(\frac{k^2}{k^2+W''_k(\phi )^2}\right).
   \label{eq:flowStrong3}
\end{align}
In terms of the dimensionless quantity $kw_t=W_k$, this reads
\begin{align}
\partial_tw_t(\phi)+w_t(\phi)=-\frac{\ln \left({1+w''_t(\phi )^2}\right)}{4
   \pi  w''_t(\phi )}.
   \label{eq:flowStrong5}
\end{align}
This equation is regular at $w''=0$ since $ \ln(1+x) = x-\frac{x^2}2 +
\frac{x^3}3 \mp \ldots$ for small $x$.  The second derivative of this equation
at the fixed point $\partial_t w_\ast=0$ is given by $(u=w_\ast''$)
\begin{align}\label{eq:flowStrong7}
	u''=&\frac{4 \pi  {u}^3 \left({u}^2+1\right)}
	{(u^ 2+1)\ln \left({{u}^2+1}\right)-2 {u}^2}\\
  &+\frac{2 {u}
   \left({u}^2-1\right){u}'^2}{\left({u}^2+1\right) 
   \left(2 {u}^2-\ln \left({{u}^2+1}\right)({u}^2+1)
   \right)}+\frac{2{u}'^2}{{u}} ,
   \nonumber
\end{align}
which agrees with Eq.~\eqref{eq:fp1} in the limit $\eta=0$.  It has the same
singularity structure as the corresponding Eq.~\eqref{eq:fixedpoint5} for the
simpler regulator used in the main text. The fourth derivative becomes
singular if
\begin{align}
	u^2=\frac12(u^2+1)\ln(u^2+1)
	\quad\Rightarrow\quad u\simeq1.9803,
\end{align}
(to be compared with the singular point $u=1$ for the simpler regulator).
Overall, we find the same types of solutions as discussed in section
\ref{sec:FixedPointStructure}. In particular, we again find oscillatory
solutions and a maximally IR-stable fixed point with one relevant direction.

A polynomial expansion of Eq.~\eqref{eq:flowStrong7} with
$u=2\lambda_t^2\phi+\sum_{n=2}^N 2n\cdot b_{2n,t}\phi^{2n-1}$ yields a system
of algebraic equations that determine the expansion coefficients of the
fixed-point potential. This system is given by Eq.~\eqref{eq:polyCoupling} without
the equation for $a^2$, with $\eta=0$ and the derivatives with respect to $t$
set equal to zero.

\label{sec:IRstableFP}

The results for the fixed-point couplings for the maximally IR-stable
fixed-point are displayed in Tab. \ref{tab:CoeffIRStable} for different
truncations. A comparison with Tab.~\ref{tab:ExponIRstabilA} reveals the
regulator-dependent variations of these nonuniversal quantities.%

We find that the coupling $\lambda_\ast$ converges quickly. Again, the
polynomial expansion provides a good approximation to the solution of the
partial differential equation for the first half period. This is displayed in
Fig.~\ref{fig:FPStrongReg}, left panel.

By contrast, the universal critical exponents $\theta^I$ at the fixed points
are much less regulator dependent. This is demonstrated in
Tab.~\ref{tab:EigsStabMatrixN16StrongReg} which should be read side by side
with Tab.~\ref{tab:EigsStabMatrixN16} for a different regulator. For all
positive critical exponents, we find a remarkable degree of universality, as
these exponents for the different regulators differ -- if at all -- by at most
10\%. The subleading negative critical exponents show larger variations and
thus require higher orders in the derivative expansion for a quantitatively
reliable prediction.

We also calculate the phase diagram with this regulator for a truncation at
$2n=10$ and compare both regulators $r_2=(k/\abs{p}-1)\theta(p^2/k^2-1)$ and
$r_2=(k^2/p^2-1)\theta(p^2/k^2-1)$ in Fig.~\ref{fig:FPStrongReg}, right
panel. The nonuniversal values for $\lam a$ at the phase transition differ
roughly by a factor of two. This clearly demonstrates that a naive comparison
of bare couplings is substantially inflicted by the regularization scheme.


\begin{thebibliography}{10}

\bibitem{Catterall:2009it}
Simon Catterall, David~B. Kaplan, and Mithat Unsal.
\newblock {Exact lattice supersymmetry}.
\newblock 2009, arXiv:0903.4881 [hep-lat].
%%CITATION = 0903.4881;%%

\bibitem{Giedt:2006pd}
Joel Giedt.
\newblock {Deconstruction and other approaches to supersymmetric lattice field
  theories}.
\newblock {\em Int. J. Mod. Phys.}, A21:3039--3094, 2006, hep-lat/0602007.
%%CITATION = HEP-LAT/0602007;%%

\bibitem{Bergner:2007pu}
Georg Bergner, Tobias Kaestner, Sebastian Uhlmann, and Andreas Wipf.
\newblock {Low-dimensional supersymmetric lattice models}.
\newblock {\em Annals Phys.}, 323:946--988, 2008, arxiv:0705.2212 [hep-lat].
%%CITATION = 0705.2212;%%

\bibitem{Kastner:2008zc}
Tobias Kaestner, Georg Bergner, Sebastian Uhlmann, Andreas Wipf, and Christian
  Wozar.
\newblock {Two-Dimensional Wess-Zumino Models at Intermediate Couplings}.
\newblock 2008, arXiv:0807.1905 [hep-lat].
%%CITATION = 0807.1905;%%

\bibitem{Aoki:2000wm}
K.~Aoki.
\newblock {Introduction to the nonperturbative renormalization group and its
  recent applications}.
\newblock {\em Int. J. Mod. Phys.}, B14:1249--1326, 2000.
%%CITATION = IMPAE,B14,1249;%%

\bibitem{Berges:2000ew}
Jurgen Berges, Nikolaos Tetradis, and Christof Wetterich.
\newblock Non-perturbative renormalization flow in quantum field theory and
  statistical physics.
\newblock {\em Phys. Rept.}, 363:223--386, 2002.
\newblock hep-ph/0005122.
%%CITATION = HEP-PH/0005122;%%

\bibitem{Litim:1998nf}
Daniel~F. Litim and Jan~M. Pawlowski.
\newblock {On gauge invariant Wilsonian flows}.
\newblock 1998.
\newblock hep-th/9901063.
%%CITATION = HEP-TH/9901063;%%

\bibitem{Pawlowski:2005xe}
Jan~M. Pawlowski.
\newblock {Aspects of the functional renormalisation group}.
\newblock {\em Annals Phys.}, 322:2831--2915, 2007, hep-th/0512261.
%%CITATION = HEP-TH/0512261;%%

\bibitem{Gies:2006wv}
Holger Gies.
\newblock {Introduction to the functional RG and applications to gauge
  theories}.
\newblock 2006, hep-ph/0611146.
%%CITATION = HEP-PH/0611146;%%

\bibitem{Sonoda:2007av}
Hidenori Sonoda.
\newblock {The Exact Renormalization Group -- renormalization theory revisited
  --}.
\newblock 2007, arXiv:0710.1662 [hep-th].
%%CITATION = 0710.1662;%%

\bibitem{Bonini:1998ec}
M.~Bonini and F.~Vian.
\newblock Wilson renormalization group for supersymmetric gauge theories and
  gauge anomalies.
\newblock {\em Nucl. Phys.}, B532:473--497, 1998, hep-th/9802196.
%%CITATION = HEP-TH/9802196;%%

\bibitem{Vian:1998kv}
F.~Vian.
\newblock Supersymmetric gauge theories in the exact renormalization group
  approach.
\newblock 1998, hep-th/9811055.
%%CITATION = HEP-TH/9811055;%%

\bibitem{Falkenberg:1998bg}
Sven Falkenberg and Bodo Geyer.
\newblock {Effective average action in N = 1 super-Yang-Mills theory}.
\newblock {\em Phys. Rev.}, D58:085004, 1998, hep-th/9802113.
%%CITATION = HEP-TH/9802113;%%

\bibitem{Arnone:2004ey}
S.~Arnone and K.~Yoshida.
\newblock {Application of exact renormalization group techniques to the
  non-perturbative study of supersymmetric field theory}.
\newblock {\em Int. J. Mod. Phys.}, B18:469--478, 2004.
%%CITATION = IMPAE,B18,469;%%

\bibitem{Arnone:2004ek}
Stefano Arnone, Francesco Guerrieri, and Kensuke Yoshida.
\newblock {N = 1* model and glueball superpotential from renormalization group
  improved perturbation theory}.
\newblock {\em JHEP}, 05:031, 2004, hep-th/0402035.
%%CITATION = HEP-TH/0402035;%%

\bibitem{Rosten:2008ih}
Oliver~J. Rosten.
\newblock {On the Renormalization of Theories of a Scalar Chiral Superfield}.
\newblock 2008, arXiv:0808.2150 [hep-th].
%%CITATION = 0808.2150;%%

\bibitem{Sonoda:2008dz}
Hidenori Sonoda and Kayhan Ulker.
\newblock {Construction of a Wilson action for the Wess-Zumino model}.
\newblock 2008, arXiv:0804.1072 [hep-th].
%%CITATION = 0804.1072;%%

\bibitem{Gies:2009az}
Holger Gies, Franziska Synatschke, and Andreas Wipf.
\newblock {Supersymmetry breaking as a quantum phase transition}.
\newblock 2009, arXiv:0906.5492 [hep-th].
%%CITATION = 0906.5492;%%


\bibitem{Synatschke:2008pv}
Franziska Synatschke, Georg Bergner, Holger Gies, and Andreas Wipf.
\newblock {Flow Equation for Supersymmetric Quantum Mechanics}.
\newblock {\em JHEP}, 03:028, 2009.
\newblock arXiv:0809.4396 [hep-th]
%%CITATION = 0809.4396;%%

\bibitem{Horikoshi:1998sw}
Atsushi Horikoshi, Ken-Ichi Aoki, Masa-aki Taniguchi, and Haruhiko Terao.
\newblock {Non-perturbative renormalization group and quantum tunnelling}.
\newblock 1998, hep-th/9812050.
%%CITATION = HEP-TH/9812050;%%

\bibitem{Weyrauch:2006aj}
M.~Weyrauch.
\newblock {Functional renormalization group and quantum tunnelling}.
\newblock {\em J. Phys.}, A39:649--666, 2006.
%%CITATION = JPAGB,A39,649;%%

\bibitem{Witten:1982df}
Edward Witten.
\newblock {Constraints on Supersymmetry Breaking}.
\newblock {\em Nucl. Phys.}, B202:253, 1982.
%%CITATION = NUPHA,B202,253;%%

\bibitem{Ranft1984166}
J.~Ranft and A.~Schiller.
\newblock {Hamiltonian Monte Carlo study of (1+1)-dimensional models with
  restricted supersymmetry on the lattice}.
\newblock {\em Physics Letters B}, 138(1-3):166 -- 170, 1984.
%%CITATION = PHLTA,B138,166;%%

\bibitem{Beccaria:2004ds}
Matteo Beccaria, Gian~Fabrizio De~Angelis, Massimo Campostrini, and Alessandra
  Feo.
\newblock {Phase diagram of the lattice Wess-Zumino model from rigorous lower
  bounds on the energy}.
\newblock {\em Phys. Rev.}, D70:035011, 2004.
%%CITATION = HEP-LAT/0405016;%%

\bibitem{Beccaria:2004pa}
Matteo Beccaria, Massimo Campostrini, and Alessandra Feo.
\newblock {Supersymmetry breaking in two dimensions: The lattice N = 1
  Wess-Zumino model}.
\newblock {\em Phys. Rev.}, D69:095010, 2004.
%%CITATION = HEP-LAT/0402007;%%

\bibitem{Golterman:1988ta}
Maarten F.~L. Golterman and Donald~N. Petcher.
\newblock {A local interactive lattice model with supersymmetry}.
\newblock {\em Nucl. Phys.}, B319:307--341, 1989.
%%CITATION = NUPHA,B319,307;%%

\bibitem{Catterall:2003ae}
Simon Catterall and Sergey Karamov.
\newblock {A lattice study of the two-dimensional Wess Zumino model}.
\newblock {\em Phys. Rev.}, D68:014503, 2003.
%%CITATION = HEP-LAT/0305002;%%

\bibitem{Beccaria:2003ba}
Matteo Beccaria and Carlo Rampino.
\newblock {World-line path integral study of supersymmetry breaking in the
  Wess-Zumino model}.
\newblock {\em Phys. Rev.}, D67:127701, 2003.
%%CITATION = HEP-LAT/0303021;%%

\bibitem{Coleman:1973jx}
Sidney~R. Coleman and Erick~J. Weinberg.
\newblock {Radiative Corrections as the Origin of Spontaneous Symmetry
  Breaking}.
\newblock {\em Phys. Rev.}, D7:1888--1910, 1973.
%%CITATION = PHRVA,D7,1888;%%

\bibitem{Murphy:1983ag}
T.~Murphy and L.~O'Raifeartaigh.
\newblock {A note on supersymmetry breaking in 1+1 dimensions}.
\newblock {\em Nucl. Phys.}, B218:484--492, 1983.
%%CITATION = NUPHA,B218,484;%%

\bibitem{Bergner:2009}
Georg Bergner.
\newblock {\em {Symmetries and the methods of quantum field theory:
  Supersymmetry on a space-time lattice}}.
\newblock PhD Thesis (University Jena), 2009.
\newblock
\url{http://www.tpi.uni-jena.de/qfphysics/thesis/georg_bergner_phd.pdf}

\bibitem{Bartels:1983wm}
Jochen Bartels and J.~B. Bronzan.
\newblock {Supersymmetry on a lattice}.
\newblock {\em Phys. Rev.}, D28:818, 1983.
%%CITATION = PHRVA,D28,818;%%

\bibitem{Wetterich:1992yh}
Christof Wetterich.
\newblock {Exact evolution equation for the effective potential}.
\newblock {\em Phys. Lett.}, B301:90--94, 1993.
%%CITATION = PHLTA,B301,90;%%

\bibitem{Neves:1998tg}
Rui Neves, Yuri Kubyshin, and Robertus Potting.
\newblock {Polchinski ERG equation and 2D scalar field theory}.
\newblock 1998, hep-th/9811151.
%%CITATION = HEP-TH/9811151;%%

\bibitem{Morris:1994jc}
Tim~R. Morris.
\newblock {The Renormalization group and two-dimensional multicritical
  effective scalar field theory}.
\newblock {\em Phys. Lett.}, B345:139--148, 1995.
%%CITATION = HEP-TH/9410141;%%

\bibitem{Weinberg:1976xy}
Steven Weinberg.
\newblock {Critical Phenomena for Field Theorists}.
\newblock Lectures presented at Int. School of Subnuclear Physics, Ettore
  Majorana, Erice, Sicily, Jul 23 - Aug 8, 1976.
  \newblock {\em Erice Subnucl.Phys.1976:1}

\bibitem{Zamolodchikov86}
A.B. Zamolodchikov.
\newblock {Conformal Symmetry and Multicritical Points in Two-Dimensional
  Quantum Field Theory}.
\newblock {\em Yad. Fiz.}, 44:529, 1986.

\bibitem{Wozar:2008jb}
Christian Wozar, Georg Bergner, Tobias Kaestner, Sebastian Uhlmann, and Andreas
  Wipf.
\newblock {Numerical Investigation of the 2D N=2 Wess-Zumino Model}.
\newblock 2008, arXiv:0809.2176 [hep-lat].
%%CITATION = 0809.2176;%%

\bibitem{strack-2009}
P.~Strack, S.~Takei, and W.~Metzner.
\newblock Anomalous scaling of fermions and order parameter fluctuations at
  quantum criticality, 2009, arXiv:0905.3894 [cond-mat.str-el].

\end{thebibliography}
\end{document}